\documentclass[11pt]{article}
\usepackage{jheppub}
\usepackage[utf8]{inputenc}
\usepackage{amsmath}
\usepackage{amssymb}
\usepackage{commath}
\usepackage[mathscr]{euscript}
\usepackage{physics}
\usepackage{braket}
\usepackage{tensor}
\usepackage{soul}
\usepackage{hyperref}
\usepackage{multirow}
\usepackage[dvipsnames]{xcolor}
\usepackage{bclogo}
\usepackage{comment}
\usepackage{dsfont}
\usepackage{cancel}
\usepackage[normalem]{ulem} 
\usepackage{threeparttablex}
\usepackage{caption}
\usepackage{subcaption}
\usepackage{graphicx}
\usepackage{color}
\usepackage{scrextend}
\usepackage{float}
\usepackage{multicol}
\usepackage{mathrsfs}

\usepackage[T1]{fontenc}
\usepackage{pifont}
\makeatletter
\def\@fpheader{\relax}
\makeatother

\numberwithin{equation}{section}

\setstcolor{red}
\begin{document}

\title{\vspace*{0.2cm} Disordered Charged Horizons}

\author[1,2]{Daniel Are\'an,}
\author[3]{Sebastian Grieninger,}
\author[1,2]{Pau G. Romeu}

\affiliation[1]{Instituto de F\'isica Te\'orica UAM/CSIC, Calle Nicol\'as Cabrera 13-15, 28049 Madrid, Spain}
\affiliation[2]{Departamento de F\'isica Te\'orica, Universidad Aut{\'o}noma de Madrid, Campus de Cantoblanco, 28049 Madrid, Spain}
\affiliation[3]{Department of Physics, University of Washington, Seattle, WA 98195}

\preprint{\begin{flushright} IFT-UAM/CSIC-25-96\\  NT@UW-25-14 \end{flushright}}

\emailAdd{daniel.arean@uam.es}
\emailAdd{segrie@uw.edu}
\emailAdd{pau.garcia@uam.es}

\vspace{1cm}

\abstract{
We construct fully backreacted charged black brane solutions with a spatially disordered chemical potential in asymptotically AdS$_3$ and AdS$_4$, providing holographic duals of strongly coupled disordered systems.
At intermediate temperatures these geometries display highly inhomogeneous horizons, though their geometric averages reproduce the clean BTZ and Reissner–Nordström solutions.
The low temperature behavior, however, differs sharply between dimensions. In AdS$_3$, inhomogeneities decay and the horizon flows to the clean charged BTZ fixed point, rendering disorder irrelevant in the infrared.
In AdS$_4$, horizon inhomogeneities persist: 
while the averaged geometry flows to the clean 
AdS$_2\times \mathbb{R}^2$ throat, the disordered horizon induces a finite residual resistivity.
These results show that disorder can qualitatively alter the IR physics of holographic metals and indicate violations of the Harris criterion in strongly coupled systems.


}

\setcounter{tocdepth}{2} 
\maketitle



\section{Introduction}
\label{sec:intro}

Disorder is ubiquitous in real materials, arising from impurities in chemical composition, defects in crystal structures, or intentional doping. No real physical system exhibits perfectly homogeneous properties.
Instead, local random deviations from the clean (homogeneous) values of quantities such as the chemical potential give rise to what is termed disorder.
In many cases these deviations average out over large scales, leaving the macroscopic behavior of the system well-described by the clean theory.
Consequently, qualitative predictions for real-world systems can often be obtained by neglecting disorder entirely~\cite{Vojta_2019,Roy:2016amv}.
However, this simplification fails for many systems and observables. A particularly notable example is that of transport coefficients: in a translationally invariant system, momentum conservation leads to divergent DC transport coefficients such as the electrical conductivity. Introducing even weak disorder breaks translational symmetry, relaxes momentum, and renders these coefficients finite. Beyond this perturbative effect, disorder can fundamentally alter the nature of electronic states, leading, for instance, to Anderson localization~\cite{Anderson:1958vr}, where strong enough disorder suppresses diffusion entirely, localizing quantum-mechanical wavefunctions. When (strong) interactions coexist with disorder, rich phenomena may emerge, such as many-body
localization~\cite{Basko_2006} or disorder-driven quantum phase transitions~\cite{PhysRevLett.105.037001}. 
Understanding this interplay is crucial, as both disorder and interactions play a decisive role in determining the low-energy physics of real materials.
(See~\cite{vojta2013phases,Vojta_2019} for a review on these phenomena).

Understanding how disorder affects the renormalization group (RG) flow is essential not only for making accurate predictions about real systems but also because disordered fixed points host a diverse range of novel phenomena with no clean counterpart, such as Griffiths phase transitions~\cite{PhysRevLett.23.17}, high-$T_c$ superconductivity, and many-body localization. In a perturbative setting, the Harris criterion~\cite{Harris:1974} provides a useful diagnostic for the effect of Gaussian disorder in the infrared (IR), classifying it as relevant, irrelevant, or marginal based on the nature of the disordered quantity, its spatial correlations, and the number of directions with disorder.
However, this criterion becomes unreliable
when the underlying clean theory is strongly coupled. More importantly,
it does not determine the low-energy physics of systems in which disorder becomes relevant in the IR.
Addressing these limitations requires non-perturbative tools, making holography a valuable framework for exploring disorder in strongly interacting systems.

Disorder in conformal and large-$N$ field theories was studied in~\cite{Aharony:2015aea}, where it was shown that it can lead to new fixed points that are not scale-invariant. 
Within the gauge/gravity duality, the authors of~\cite{Hartnoll:2008hs} studied momentum relaxation due to magnetic and electric impurities. 
The study of holographic theories with disordered charge density was initiated
in~\cite{Adams:2011rj,Adams:2012yi}.
These works considered a perturbatively small disordered chemical potential in Einstein-Maxwell theory.
This approach was further pursued in~\cite{Garcia-Garcia:2015crx} (see also~\cite{Andrade:2017lsc}) where
both the Harris-irrelevant and Harris-marginal settings were studied perturbatively for disorder about charge neutrality. In~\cite{OKeeffe:2015qma} the analysis of disorder about a charged black hole was started, also through a perturbative approach.
Focusing on transport properties of theories dual to disordered black holes,~\cite{Grozdanov:2015qia}
proved a lower bound on the electrical conductivity thereby excluding the possibility of realizing many-body localization in holographic theories with connected horizons. This was extended in~\cite{Grozdanov:2015djs} to thermal transport.

The explicit implementation of disorder in holography was initiated in~\cite{Arean:2013mta} for superfluid systems in the probe limit.
Going beyond the probe limit, \cite{Hartnoll:2014cua,Hartnoll:2015faa,Hartnoll:2015rza} constructed holographic duals to theories at zero and finite temperature 
in which disorder was introduced via the source of a scalar operator.
By tuning the dimension of the disordered coupling to make it Harris-marginal, those works found that the theory flows to a Lifshitz-scaling fixed point in the IR.
The emergence of IR fixed points in disordered theories was further
analyzed in~\cite{Ganesan:2020wzm,Ganesan:2021gun}, which carefully studied the RG flow in holographic field theories with quenched disorder.
Finally,~\cite{Huang:2023ihu} constructed holographic quantum critical points arising from the introduction of a disordered scalar operator whose dimension is perturbatively close to Harris-marginal.

In this work, we employ the gauge/gravity duality to construct fully backreacted, asymptotically AdS black hole geometries with disordered $U(1)$ electric fields. These setups are dual to strongly coupled matter with a Harris-relevant disordered chemical potential. The resulting geometries exhibit highly inhomogeneous horizons, reflecting the microscopic disorder of the dual theory. By studying these systems at finite charge density, we address two key questions.
First, we determine how a disordered electric field modifies the geometry of the planar horizon.
Second, by analyzing the effect of disorder in the IR, we 
study the low-energy dynamics of the disordered system and quantify its eventual departure from the quantum critical point governing the ground state of the clean system.
Indeed, in absence of disorder the low temperature near-horizon geometry features an AdS$_2$ factor.
Thus, it corresponds to a quantum critical phase with an emergent time-scaling symmetry, termed the semi-local quantum critical phase~\cite{Liu:2009dm,Cubrovic:2009ye,Faulkner:2009wj,Iqbal:2011in}.

This paper is organized as follows. In Section~\ref{sec:disorder} 
we briefly review quenched disorder in quantum field theory and the Harris criterion.
We then outline the numerical implementation of disorder we will employ in this work.
Section ~\ref{sec:charged_horizons} introduces the holographic model and discusses how it accommodates clean asymptotically AdS$_3$ and AdS$_4$ charged black brane geometries.
We also present the ansatzes and boundary conditions that allow for the introduction of disorder,
and describe how DC transport coefficients can be extracted from horizon data of the disordered solutions.
Our main results are presented in Sections~\ref{sec:ads4} and~\ref{sec:ads3}, where we analyze the disordered geometries in AdS$_4$ and AdS$_3$, respectively. We characterize their low temperature behavior and determine their DC conductivities.
We conclude in Section~\ref{sec:conclusions} with a discussion of our results and directions for future research.


\section{Disorder}
\label{sec:disorder}
Disorder in quantum field theories can be introduced  through a spatially inhomogeneous coupling whose variation follows a random distribution.
This approach, in which the spatially dependent coupling is taken to be time-independent, is called quenched disorder.
Naturally, all expectation values depend on the particular realization of the disordered coupling. 
Because of this realization dependence, the physically meaningful observables are disorder-averaged quantities.
These are computed by performing a path integral over the coupling, with a probability functional that encodes the statistical properties of disorder.

In the case of Gaussian disorder, one can employ the replica trick to express disorder-averaged expectation values in terms of those of an homogeneous theory with nonlocal couplings~\cite{Aharony:2018}, where the disorder strength appears as an effective (homogeneous) coupling constant. This framework allows us to analyze how the clean theory responds to the introduction of disorder. By applying standard power-counting arguments to the homogeneous nonlocal coupling, we recover the Harris criterion~\cite{Harris:1974, Vojta_2019}. 
This criterion determines the fate of disorder under the flow to the infrared.
If the mass dimension of the disorder strength is positive, negative, or 0,
disorder is relevant, irrelevant, or marginal, respectively.
Therefore, when an operator of dimension $\Delta$ is coupled to a source with Gaussian (white) noise, disorder becomes relevant if the following inequality holds: 
\begin{equation}
    \Delta < d-\frac{n}{2} \, , 
\end{equation}
where $d$ is the spacetime dimension of the field theory and $n$ is the number of
disordered dimensions.
In this paper, we introduce disorder through a noisy chemical potential. 
The chemical potential couples to a conserved current of dimension $\Delta_J = d-1$. 
Therefore, for disorder to be relevant,
$n$ must be strictly less than 2.
In most physical systems quenched disorder is present in all spatial directions.
This type of disorder is known as isotropic quenched disorder
and is characterized by $n=d-1$.
This singles out $d=2$ as the only scenario in which isotropic quenched disorder sourced by a chemical potential is relevant. 
For an arbitrary spacetime dimension $d$, we must restrict ourselves to $n=1$ to obtain  Gaussian Harris-relevant disorder.

In holography, a disordered charged system is realized by means of an inhomogeneous source of the temporal component of the gauge field dual to the chemical potential $\mu$.
The relevant observables are the expectation values of operators
averaged over different choices for the inhomogeneous source $\mu(x)$.
Different choices of sets to average over define different types of disorder. 
We choose an ensemble $\{\mu(x)\}$ that generates local white Gaussian noise, 
namely maximal disorder with no correlations between arbitrarily nearby points.
The disorder-averaged 1- and 2-point functions for $\mu(x)$ define this disorder
\begin{align}
\langle\mu(x)\rangle_D = \mu_0\,,\qquad
\langle\mu(x)\,\mu(y)\rangle_D= \mu_0^2+ V^2\, \delta(x-y)\,,
\label{eq:disdistro}
\end{align}
where $V$ is the disorder strength.
Since disorder is introduced only along one spacelike direction,
$\left[\delta(x)\right]=\left[\mu_0 \right]=1$. 
Therefore, 
$\left[V\right]=1/2$, 
and disorder is Harris-relevant in this case.

A numerical representation of disordered sources fulfilling~\eqref{eq:disdistro} follows~\cite{Arean:2013mta,Hartnoll:2014cua,Garcia-Garcia:2015crx}.
The inhomogeneous chemical potential is taken as
\begin{align}
    \mu(x)= \mu_0\left[1 +w\sum_{n=1}^{N}
    \cos\left(n\,k_0/N x +\delta_n\right)\right],
    \label{eq:dismu}
\end{align}
where each $\delta_n$ is a random number drawn uniformly from $[0,2\pi]$.
One can check that~\eqref{eq:dismu}
realizes the local Gaussian noise~\eqref{eq:disdistro} in the limit $N\to\infty$ ~\cite{Arean:2014oaa,Hartnoll:2014cua,Garcia-Garcia:2015crx}, with the disorder average $\langle\rangle_D$ of a quantity $g$ being defined as
\begin{equation}
    \langle g \rangle_D = \int\prod_{n=1}^{N}{d\delta_n\over 2\pi}\, g\,.
    \label{eq:disavdef}
\end{equation}
We use (\ref{eq:disavdef}) to compute the disorder averages in (\ref{eq:disdistro}), replacing the delta function
by its avatar with $N$ and $k_0$ finite 
\begin{equation}
    \delta_{(N,\,k_0)} (x-y) \equiv  \frac{k_0}{\pi N}\sum_{n=1}^{N} \cos\left( k_0 \frac{n}{N} (x-y)\right) .
    \label{eq:discdelt}
\end{equation}
For modes with a wavevector that is an integer multiple of $k_0/N$ but less than $k_0$,
$\delta_{(N,k_0)}$
behaves as a delta function when integrated over $[ -\pi N/k_0,\,\pi N/k_0]$.
Thus, the dimensionful disorder strength $V$ can be expressed
in terms of the dimensionless parameter $w$ as
\begin{equation}
V= w \,  \sqrt{ \frac{\pi\mu_0}{2} \, \frac{N }{k_0 / \mu_0}} \, .
\end{equation}

The numerical implementation of disorder~\eqref{eq:dismu} at finite $N$ introduces two scales.
The smallest wavevector, $k_0/N$, sets the periodicity of $\mu(x)$ and thus fixes the length of our system as $L_x=2\pi N/k_0$. The largest wavevector, $k_0$, acts as an ultraviolet (UV) cutoff of the disorder distribution; \textit{e.g.}, probing distances smaller than $1/k_0$ reveals
oscillations of frequency $k_0$ around the delta function distribution~\eqref{eq:discdelt}.

As a consequence of the UV and IR cutoffs, the two-point function in \eqref{eq:disavdef} becomes a discrete avatar of the delta function \cite{Arean:2014oaa}. It can only be distinguished from an actual delta by probing energies bigger than the disorder UV cutoff.
To recover the usual delta function one would first take the UV cutoff to infinity,
$k_0 \to \infty$, 
and then the infinite size limit $ N/k_0 \to \infty$.
However, this limit is not physically sensible.
From the gravitational perspective, 
sending the UV cutoff to infinity would destroy the asymptotically AdS region of spacetime, making the holographic dictionary ill-defined. 
From the dual field-theory perspective,
taking $k_0 \to \infty$ makes the system disordered down to arbitrarily short distances,
which is atypical of real disordered systems.
Indeed, consider a material with dilute impurities: at length scales much smaller than the average impurity spacing, the system is effectively clean.
Therefore, the physically meaningful limit is to keep the UV cutoff fixed and take $N \to \infty$. 

All in all, taking into account that our holographic model is conformal in the UV, and thus its dynamics is controlled by
dimensionless ratios, 
the IR and UV cutoffs discussed above restrict the range of temperatures of our disordered system to
\begin{align}
   &k_{\rm IR}<  a{T\over\mu_0}< k_{\rm UV}\,,\nonumber\\
   \text{with}\quad
   &k_{\rm IR}={k_0\over N\mu_0}\,,\quad
   k_{\rm UV}={k_0\over\mu_0}\,.
   \label{eq:discutoffs}
\end{align}
The extra factor of $a$ 
in front of the temperature follows from the relation between the horizon radius and temperature. It takes the value $a=4\pi/3$ for AdS$_4$, and $a=2\pi$ for AdS$_3$.


\section{Disordered charged horizons}
\label{sec:charged_horizons}
In this section, we first introduce the theoretical setup that will allow us to study
strongly coupled systems in the presence of disorder.
We then specialize the framework to asymptotic AdS$_4$ and AdS$_3$ geometries, corresponding to disordered 2+1 and 1+1 dimensional theories. Finally, we present expressions for the DC conductivities of the dual theories in terms of the geometry of the inhomogeneous horizons. 

In order to construct duals to strongly coupled field theories at nonzero charge density of a $U(1)$ global symmetry we consider solutions of the Einstein-Maxwell theory
\begin{align}
S=\int d^{d+1}x\, \sqrt{-g}\, \left(R-2 \Lambda -{1\over4} F^2
\right).
\label{eq:einsmaxwellaction}
\end{align}
We  study the cases $d=2$ and $d=3$ corresponding to black hole geometries asymptotic to AdS$_3$ and AdS$_4$, respectively,
with $\Lambda = -d(d-1)/2 L^2$;
henceforth we set $L=1$ unless explicitly stated.  
The resulting equations of motion are given in Appendix~\ref{app:eomsbcs}, where we also discuss the relevant boundary conditions. 

\subsection{AdS$_4$. Disordered AdS-Reissner-Nordstr\"om}
\label{ssec:AdS4RN}

We  first study inhomogeneous charged black holes in AdS$_4$.
These are dual to strongly coupled systems at nonzero charge density where the chemical potential is disordered along one spatial direction.

The homogeneous phase is given by the planar AdS$_4$-Reissner–Nordstr\"om (RN) black hole geometry
\begin{align}
    &ds^2={1\over z^2}\left[
    -G\,dt^2+{dz^2\over G}+dx^2+dy^2
    \right]\,,\qquad
G=(1-z)\left(1+z+z^2-{\bar\mu^2\over4}\,z^3\right)\,, \nonumber\\ &A_t=\bar\mu\,(1-z)\,.
\label{eq:homads4}
\end{align}
We have made use of the scaling invariance 
$(x^\mu,z)\to\lambda(x^\mu,z)$ to set the horizon at $z=1$.
Its temperature is given by\footnote{Technically $T$ and $\bar\mu$ are the temperature and chemical potential in units of the horizon radius. Since only dimensionless ratios are physically sensible, we 
work in terms of the dimensionless parameters $\bar\mu$ and $T$.}
\begin{equation}
T={12-\bar\mu^2\over16\pi}\,.
\label{eq:tempads4}
\end{equation}
As is obvious from the form of $A_t$, $\bar\mu$ corresponds to the chemical potential of the dual theory, and one can easily check that it also fixes the charge density $\langle J^t\rangle=\bar\mu$.

We next consider a disordered chemical potential given by eq.~\eqref{eq:dismu}.
The resulting geometry depends on both $x$ and $z$.
We assume translational invariance along the remaining spatial direction
$y$ and thus take the following  static ansatz~\cite{Donos:2014yya}
\begin{align}
\label{eq:ads4ansatz}
&ds^2= {1\over z^2}\left[
-h_1\,G\,dt^2+{h_2\over G}\,dz^2+h_3\left(dx+h_5\,dz\right)^2
+h_4\,dy^2\right]\,,\qquad A_t = A_t(z,x)\,,
\end{align}
where $G$ is given in~\eqref{eq:homads4}.

Our ansatz~\eqref{eq:ads4ansatz} features six undetermined functions of $(z,x)$: $h_1$, $h_2$, $h_3$, $h_4$, $h_5$, and $A_t$. It reduces to the homogeneous AdS$_4$-RN black hole~\eqref{eq:homads4} when
$h_1=h_2=h_3=h_4=1$, $h_5=0$, and $A_t=\bar\mu(1-z)$.
As explained in~\cite{Horowitz:2012ky,Donos:2014yya},
we have not used diffeomorphism invariance along $x$ to further restrict our metric ansatz. This allows us to employ the DeTurck trick~\cite{Headrick:2009pv} and obtain a set of six second order partial differential equations (PDE), as discussed in Appendix~\ref{app:eomsbcsads4}.

To solve the PDEs numerically, we have to impose boundary conditions at the IR horizon $z=1$ and the UV boundary $z=0$.
At $z=1$, we require the existence of a regular horizon
which results in the boundary conditions detailed in
appendix~\ref{app:eomsbcsads3}.
It is straightforward to check that the temperature of the horizon is given
by~\eqref{eq:tempads4}. 
In the UV,  we apply the holographic dictionary and switch on a
disordered chemical potential via
$A_t(0,x)= \mu(x)$ with $\mu(x)$ given in~\eqref{eq:dismu}.
Notice that the parameter $\bar\mu$ that sets the temperature via~\eqref{eq:tempads4} no longer corresponds to the chemical potential.
As for the metric functions we demand that they asymptote to their AdS$_4$ values
\begin{equation}
h_1(0,x)=h_2(0,x)=h_3(0,x)=h_4(0,x)=1\,,\; h_5(0,x)=0\,.
\label{eq:uvbcsads4}
\end{equation}

Our numerical simulations generate inhomogeneous geometries that asymptote to AdS$_4$,
with a charged horizon that is highly inhomogeneous along $x$. 
For a given disorder realization, namely a particular choice of $N$, $k_0$ and random phases $\delta_n$, we obtain a family of solutions parametrized by the dimensionless ratio of temperature over the average of the chemical potential, $T/\mu_0$.

We present the results of our numerical simulations of disordered
AdS$_4$ black holes in Section~\ref{sec:ads4}, where we also discuss their transport properties and low temperature behavior.

\subsection{AdS$_3$. Disordered BTZ}
We now consider disordered charged black holes in asymptotic
AdS$_3$. These geometries are dual to strongly coupled theories in $1+1$ dimensions.
The geometry of the homogeneous phase is given by the charged BTZ black brane
\begin{align}
    &ds^2={1\over z^2}
    \left(-f(z)\,dt^2+{dz^2\over f(z)}+dx^2
    \right)\,,\quad f(z)=1-z^2+{\bar{\mu}^2 z^2\over2}
    \log z\,,\nonumber\\
    &A_t = \bar{\mu}\,\log z\,,
    \label{eq:BTZBH}
\end{align}
where as for AdS$_4$ we have set the radius of the horizon to
$z_h=1$.
Note that $A_t$ diverges logarithmically toward the boundary, so holographic renormalization is required. It was carried out in \cite{Jensen:2010em}, which showed that in order to work in an ensemble with
fixed $\mu$, one must add (local) finite counterterms. These are boundary terms that do not alter the equations of motion, and the procedure leads to a boundary theory where $\mu$ is the source of a dynamical gauge field. (We refer to  \cite{Jensen:2010em,Faulkner:2012gt} for further details).
The temperature of the horizon is
\begin{equation}
T={4-\bar\mu^2\over8\pi}\,.
\end{equation}

As above, we study the inhomogeneous geometries 
sourced by a disordered chemical
potential $\mu=\mu(x)$ given in~\eqref{eq:dismu}.
We adopt the ansatz
\begin{equation}
    ds^2= \left(\frac{L}{z}\right)^2 \left( -H_1\,f(z)\, dt^2+ \frac{H_2\,dz^2}{f(z)}+ H_3\,(dx+H_4\, dz)^2  \right)\,,
    \quad A = \phi\, dt\,,
\label{eq:ads3ansatz}
\end{equation}
where $f(z)$ is given in~\eqref{eq:BTZBH}. As before, we have not completely fixed diffeomorphism invariance along $x$, which allows us  to employ the DeTurck trick and obtain a system of elliptic PDEs~\cite{Dias:2015nua}.
Indeed, as detailed in Appendix~\ref{app:eomsbcs}, the equations of motion reduce to
five coupled second order PDEs for the five unknown functions of $(z,x)$ that make up our ansatz: $H_1$, $H_2$, $H_3$, $H_4$, and $\phi$.

To obtain numerical solutions, we proceed as in the AdS$_4$ case.
At the UV boundary, $z=0$, we impose AdS$_3$ asymptotics
\begin{equation}
\label{eq:ads3metricbc}
H_1=H_2=H_3=1\,, H_4=0\,.
\end{equation}
We set the disordered chemical potential via the logarithmic asymptotics\footnote{This logarithmic UV divergence requires a subtraction in the equations of motion as we discuss in Appendix~\ref{app:eomsbcs}.}
of the gauge field,
$\lim_{z\to0}\phi(x)=\mu(x)$ with $\mu(x)$ given by~\eqref{eq:dismu}. In the IR we impose regularity at the black hole horizon located at
$z=1$, obtaining the boundary conditions listed in Appendix~\ref{app:eomsbcs}. 

Our numerical simulations produce AdS$_3$ charged inhomogeneous black brane geometries,
which we analyze in Section~\ref{sec:ads3}.

\subsection{DC Transport}
\label{ssec:DCtransport}

Donos and Gauntlett~\cite{Donos:2014uba,Donos:2014cya} showed that the direct current (DC) thermoelectric conductivities of holographic theories with momentum dissipation
can be derived from horizon data of the gravity dual.
 This method, extended to inhomogeneous geometries in  \cite{Donos:2014yya,Rangamani:2015hka}, allows us to 
express the DC transport coefficients of our disordered solutions solely in terms of the horizon geometry. The procedure defines radial invariants using a Killing vector and
applying Maxwell's and Einstein's equations.

For any Killing vector $ k $, we have:
\begin{equation}
\nabla_\alpha  \nabla^\beta  k^\alpha = R^\beta_\alpha k^\alpha\,.
\label{eq:KillingRicciId}
\end{equation}
Using Einstein's equations\footnote{Numerically, we are not solving Einstein's equation but the Einstein-DeTurck equations. Following the same procedure, we see that the DeTurck term adds an extra piece to $G$, which does not contribute to $Q$. Then, corrections to the conductivity due to the DeTurck-ing come at order $\xi^2$. } together with $\mathcal{L}_k  F =0$, which follows because $k$ is an isometry of our solution, we can rewrite \eqref{eq:KillingRicciId} as 
\begin{equation} 
    \nabla_\mu G^{\mu \nu} = - \frac{2\Lambda}{D-2} k^\nu \ , 
    \label{eq:G_conservation}
\end{equation}
where  $\Lambda$ is the cosmological constant and $G^{\mu \nu}$ an antisymmetric tensor defined as
\begin{equation}
    G_{\mu \nu} = \nabla_{\mu} k_\nu + \frac{1}{2}  \left( \frac{1}{D-2} \psi -\theta  \right) F_{\mu \nu} +\frac{1}{2 \left( D-2 \right) } \left( k_\mu F_{\nu \alpha}- k_\nu F_{\mu \alpha} \right) A^\alpha \ ,
\end{equation}
where, following~\cite{Donos:2014yya}, we have defined
$k^\mu F_{\mu \nu} = \nabla_ \nu \theta$, and
$k^\mu A_\mu = \psi - \theta$.
Note that $\theta$ is gauge invariant, while $\psi$ shifts as
$\psi \ \to \ \psi + \iota_k d\lambda$ under
$A \ \to \ A +d\lambda$. Thus $G$ itself is not gauge invariant,
but \eqref{eq:G_conservation} is.
Since we focus on static solutions,
we have 
$k \propto \partial_t$ and $\partial_t (\sqrt{-g}\,G^{\mu \nu})=0$. Hence
using the identity \eqref{eq:G_conservation} we find that $\sqrt{-g}\,G^{zx}$ must be a constant. We can apply the same logic to the Maxwell equations to show that $\sqrt{-g}\,F^{xz}$
must also be a constant. So we define
\begin{equation}
\begin{aligned}
        Q \equiv \sqrt{-g}\,  G^{x z} \,,\quad
        J \equiv \sqrt{-g}\,  F^{x z} \, .
\end{aligned}
\label{eq:constantsQJ}
\end{equation}
It is easy to check from the UV asymptotics of the fields that $J$ corresponds to the current along the $x$ direction in the dual theory.
Similarly, upon defining the energy-momentum tensor in the dual field theory and expanding $Q$ near the boundary one finds that it matches the heat current.

The invariants~\eqref{eq:constantsQJ} are trivial in our background.
Introducing time-dependent perturbations, however, leads to nontrivial relations that allow us to determine the DC conductivities of the dual theory.
We therefore add linear perturbations in time to the gauge field and metric as follows
\begin{equation}
\begin{aligned}
    \delta  ds^2 & = \left(\frac{L}{z}\right)^2 
    \left[ \delta g_{\mu \nu}\, dx^\mu dx^\nu  + 2 \, g_{tt}(z,x) \, \zeta\, t \ dx\,dt \right] , \\ 
    \delta  A  & =  \delta a_{\mu }\, dx^\mu  - \left(E - A_t(z,x)\,
    \zeta\right) t \,  dx \, .
\end{aligned}
\label{eq:perturbationsDC}
\end{equation}
In the dual field theory, these perturbations correspond to applying a constant electric field $E$ and a temperature gradient $\zeta$ both along the $x$ direction.
Next, we expand these perturbations about the horizon and impose infalling boundary conditions. 
Substituting the resulting expansions in the conserved quantities~\eqref{eq:constantsQJ} and expanding them around the horizon
allows us to express $J$ and $Q$ in terms of $E$ and $\zeta$ and 
background horizon data.
Recalling the definition of the thermoelectric conductivities
\begin{equation}
    \begin{pmatrix}
        J \\
        Q
    \end{pmatrix} 
    =  
    \begin{pmatrix}
        \sigma & \alpha T \\
        \Bar{\alpha}T & \Bar{\kappa} T
    \end{pmatrix}
    \begin{pmatrix}
        E \\
        \zeta
    \end{pmatrix}\,,
\label{eq:thermoelectricDef}
\end{equation}
we can express the DC thermoelectric conductivities $\sigma$, $\alpha$, and $\kappa$ in terms of the background functions evaluated at the horizon.
Here, $\sigma$ and $\kappa$ denote the electric and thermal conductivity, respectively. 
In our case, the thermoelectric conductivities $\alpha$ and $\bar{\alpha}$ must 
coincide due to the time-reversal invariance of the UV CFT, and we indeed find this to be the case. In the following subsections we specify the results for AdS$_4$ and AdS$_3$.

\subsubsection{AdS$_4$}
\label{ssec:conductivitiesAdS4}
Let us apply the procedure above to the case of asymptotic AdS$_4$ black holes presented in
Section~\ref{ssec:AdS4RN}.
Imposing regularity of the perturbations \eqref{eq:perturbationsDC} at the horizon leads to the following expansions
\begin{equation}
\begin{aligned}
    \delta g_{tx} &=   \zeta \, \frac{h_1 G(z)}{G'(1)} \, \log(1-z) + \frac{\sqrt{h_3 h_4}}{h_4} \, \delta g_{tx}^{(0)} \, 
    ,  \\ 
    \delta a_x  &= \frac{E}{G'(1)}\, \log(1-z)  + \delta a_x^{(0)} \, ,\\
     \delta a_z &= \frac{1}{G(z)} \delta a_z^{(0)} \, , \quad   \delta a_t  =\delta a_z^{(0)}\,.   
\end{aligned}
\end{equation}
At leading order, expanding the invariants~\eqref{eq:constantsQJ} near the horizon
we find
\begin{subequations}\label{eq:JQAdS4}
\begin{align}
    J &= \sqrt{\frac{h_4}{h_3}} \left( E - \partial_x \delta a_z^{(0)}  \right) - \frac{\partial_z A_t}{h_1}\,\delta g_{t x}^{(0)} \,,\label{eq:JAdS4}\\ 
    Q &= G'(1)\, \delta g_{t x}^{(0)} \,,\label{eq:QAdS4}
\end{align}
\end{subequations}
where all functions are evaluated at the horizon $z=1$.
At the next order in the horizon expansion, only the equation for $Q$ in~\eqref{eq:constantsQJ}
yields a nonzero constraint. It reads
\begin{equation}
    \partial_x (...) =  G'(1) \frac{\partial_z A_t}{h_1} \left( E - \partial_x \delta a_z^{(0)}  \right) -  G'(1)^2 \zeta + G'(1) \delta g_{t x}^{(0)} \, \frac{1}{\sqrt{h_4 h_3}} \left( \frac{\partial_x h_4}{h_4} \right)^2 .
    \label{eq:pertconstAdS4}
\end{equation}
Here the ellipsis denotes a lengthy expression that, as we will see, does not contribute to our result; and, as above, all functions are evaluated at the horizon.
Note that \eqref{eq:QAdS4} fixes $\delta g_{t x}$ at the horizon in terms of $Q$.
Substituting this into \eqref{eq:JAdS4} and \eqref{eq:pertconstAdS4}, and integrating along the horizon, we use the periodicity of the system to discard the total derivative on the left-hand side of \eqref{eq:pertconstAdS4}.
This yields two relations expressing 
$J$ and $Q$ in terms of $E$ and $\zeta$, and background horizon data.
The transport coefficients can then be read off using the definition
\eqref{eq:thermoelectricDef}.
For convenience, we define the following horizon averages
\begin{equation}
\begin{aligned}
\label{eq:ads4xaverage}
    \braket{ \mathbf{X} }  = \frac{1}{Z} \int dx\,\frac{\sqrt{h_3\,h_4}}{h_4}\, \mathbf{X}
    \quad\text{with}\quad  Z= \int dx\,\frac{\sqrt{h_3\, h_4}}{h_4} \, .
\end{aligned}
\end{equation}
With these definitions, the thermoelectric conductivities can be written as
\begin{subequations}
\label{eq:sigmaDCAdS4}
\begin{align}
  & \sigma   =   \frac{1}{Z} \left( 1 +  \frac{\braket{\rho}^2}{\braket{\rho^2}-\braket{\rho}^2 + \braket{\Upsilon^2} }\right) ,\quad
    \alpha  =   \frac{4 \pi }{Z} \left(\frac{\braket{\rho}}
    {\braket{\rho^2}-\braket{\rho}^2 + \braket{\Upsilon^2} }\right) ,  
    \label{eq:sigmalphaDCAdS4}\\
  &\kappa  =   \frac{\left( 4 \pi \right)^2}{Z}\,T  \left(\frac{1}{\braket{\rho^2}-\braket{\rho}^2 + \braket{\Upsilon^2} }\right) ,
  \label{eq:kappaDCAdS4}
\end{align}
\end{subequations}
where we have defined
\begin{equation}
\label{eq:rho_upsil_def}
\rho = \frac{A_t'}{h_1}\quad \text{and} \quad 
\Upsilon^2 =  \frac{1}{h_3} \left( \frac{\partial_x h_4}{h_4} \right)^2  \,,
\end{equation}
with all functions evaluated at the horizon.
The structure of the denominators in~\eqref{eq:sigmaDCAdS4} will be crucial for the behavior of the conductivities.
Notice that $\rho$ corresponds to the electric field at the horizon while, as we discuss below,
$\braket{\Upsilon^2}$ is a measure of the inhomogeneity of the horizon geometry. 

In the large temperature limit 
analytic expressions for the DC conductivities can be obtained.
As $T \to \infty$ the backreaction of the matter fields vanishes. Then, approximating the geometry by that of 
a homogeneous, uncharged black brane with $A \approx \mu(x) (1- z/z_h)\,dt$,
one arrives at
\begin{equation}
\braket{\rho} = \frac{3 \mu_0 }{4\pi T }\, , \quad
\braket{\rho^2} = \left(\frac{3\mu_0 }{4\pi T } \right)^2 + \frac{9 V^2 k_0}{8 \pi^3 T^2} \, , \quad
\braket{\Upsilon^2} = 0 \,,\quad
Z=1\, ,
\end{equation}
and thus the conductivities~\eqref{eq:sigmaDCAdS4} become
\begin{equation}
\begin{aligned}
\sigma_\infty  =  1 +  \frac{\pi \mu_0^2}{V^2 k_0 }\,,\quad
    \alpha_\infty   =    \frac{\left(4 \pi \right)^2 T}{3}\frac{\pi \mu_0}{V^2 k_0 }   \,, \quad
    \kappa_\infty   &= \frac{\left(4 \pi \right)^4 T^3}{9} \frac{\pi }{V^2 k_0 } \, .  \\
\end{aligned}
\label{eq:cond_inf}
\end{equation}

\subsubsection{AdS$_3$}
\label{ssec:conductivitiesAdS3}
The DC conductivities for the asymptotically AdS$_3$
disordered geometries can be computed by following the same procedure.
The geometry is described by the ansatz~\eqref{eq:ads3ansatz}.
In this ansatz, the invariants~\eqref{eq:constantsQJ} take
a form analogous to their AdS$_4$ counterparts. At leading order towards
the horizon they read
\begin{equation}
    J = \frac{1}{\sqrt{H_3}} \left( E - \partial_x \delta a_z^{(0)}  \right) - \frac{\partial_z A_t}{H_1} \delta g_{t x}^{(0)} \,,\quad
    Q = f'(1) \delta g_{t x}^{(0)} \, ,
\label{eq:JQAdS3}
\end{equation}
with all functions evaluated at the horizon. 
We have redefined the perturbations to ensure regularity at the horizon, which yields the  expansions
\begin{equation}
\begin{aligned}
    \delta g_{tx} &=   \zeta \, \frac{H_1 f(z)}{f'(1)} \, \log(1-z) + \sqrt{H_3} \, \delta g_{tx}^{(0)} \, ,  \\ 
    \delta a_x  &= \frac{E}{f'(1)}\, \log(1-z)  + \delta a_x^{(0)} \, ,\\
     \delta a_z &= \frac{1}{f(z)} \delta a_z^{(0)} \, , \quad   \delta a_t  =\delta a_z^{(0)}\, .
\end{aligned}
\end{equation}
Expanding $Q$ to the next order yields the following nontrivial constraint
\begin{equation}
    \partial_x (...) =  f'(1) \frac{\partial_z A_t}{H_1} \left( E - \partial_x \delta a_z^{(0)}  \right) +  f'(1)^2 \zeta  \, ,
    \label{eq:pertconstAdS3}
\end{equation}
where, as before, the ellipsis denotes a lengthy expression that vanishes upon integration along the horizon and therefore does not contribute to the final result.
Notably, this constraint does not include a term as the last one
in~\eqref{eq:pertconstAdS4}, which introduced a dependence on the
inhomogeneities of the horizon along the extra spatial direction $y$
(absent in AdS$_3$).
Following the same steps as in the AdS$_4$ case, upon integration along
the horizon, \eqref{eq:JQAdS3} and~\eqref{eq:pertconstAdS3}
yield two equalities from which the thermoelectric transport coefficients can be extracted.
In line with section \ref{ssec:conductivitiesAdS4} we define 
\begin{equation}
\begin{aligned}
    \braket{ \mathbf{X} }  = \frac{1}{Z} \int \sqrt{H_3}\,\mathbf{X}  \quad \text{with}\quad Z= \int \sqrt{H_3 } \, .
\end{aligned}
\end{equation}
The DC transport coefficients then take the form
\begin{subequations}
\label{eq:sigmaDCAdS3}
\begin{align}
&\sigma =   \frac{1}{Z} \left( 1 +  \frac{\braket{\rho}^2}{\braket{\rho^2}-\braket{\rho}^2 }\right) ,  \quad
\alpha =   \frac{4 \pi }{Z} \left(\frac{\braket{\rho}}{\braket{\rho^2}-\braket{\rho}^2  }\right) ,  \\
&\kappa  =   \frac{\left( 4 \pi \right)^2 T }{Z} \left(\frac{1}{\braket{\rho^2}-\braket{\rho}^2  }\right) .
\end{align}
\end{subequations}
These expressions are formally similar to those for AdS$_4$ in
eqs.\eqref{eq:sigmaDCAdS4}, but with one important difference: in the AdS$_3$ case the only dependence enters through $Z$, and there is no analogous term to $\Upsilon$.


\section{Disordered black holes in AdS$_4$}
\label{sec:ads4}

In this section, we  present our results for disordered black brane geometries asymptotic to AdS$_4$.
We first focus on intermediate temperatures, finding highly inhomogeneous horizons.
Next we analyze the low temperature disordered geometries: 
at very low temperature our solutions feature a disordered near-horizon geometry. 
The introduction of disorder drives the ground state of the theory away from the AdS$_2\times \mathbb{R}^2$ near-horizon geometry obtained for a homogeneous chemical potential.
We also present results for the DC transport coefficients. In particular, we find that the  disordered low temperature horizon gives rise to a residual resistivity in the dual theory.

\begin{figure}[h!]
        \centering
        \begin{subfigure}{0.49\linewidth}
            \centering
            \includegraphics[width=\linewidth]{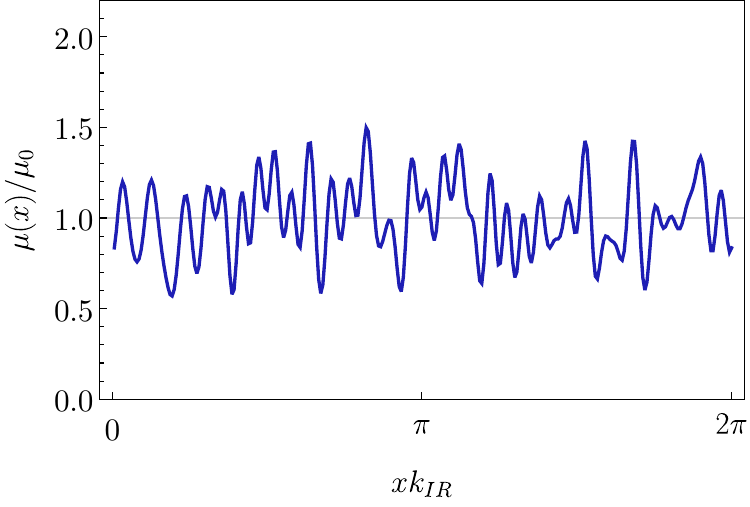}
        \end{subfigure}
        \begin{subfigure}{0.49\linewidth}
            \centering
            \includegraphics[width=\linewidth]{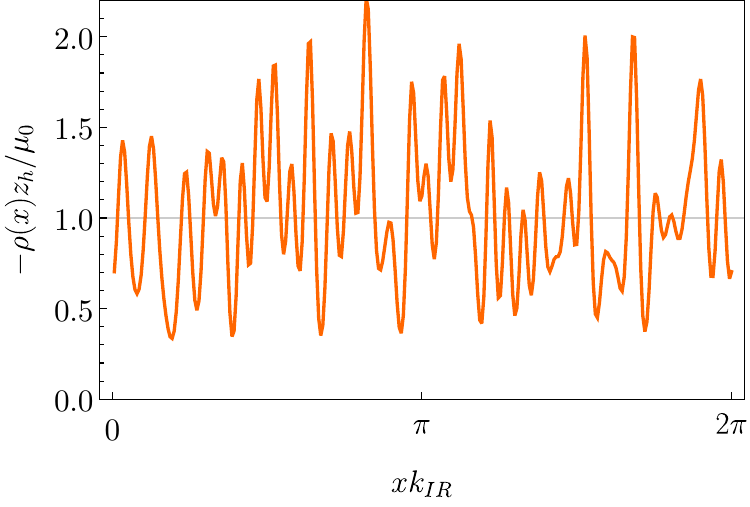}
        \end{subfigure}
        \caption{Disordered chemical potential. Left: Chemical potential $\mu(x)$. Right: Associated charge density $\rho(x)$. Both plots correspond to the same noise realization with $T/\mu_0= 0.06$, $V / \sqrt{\mu_0} = 2.5  $, $N=40$, and $k_0/\mu_0=0.06$.}
        \label{fig:dismurhoAdS4}
\end{figure}

\begin{figure}
    \centering
    \includegraphics[width=0.8\linewidth]{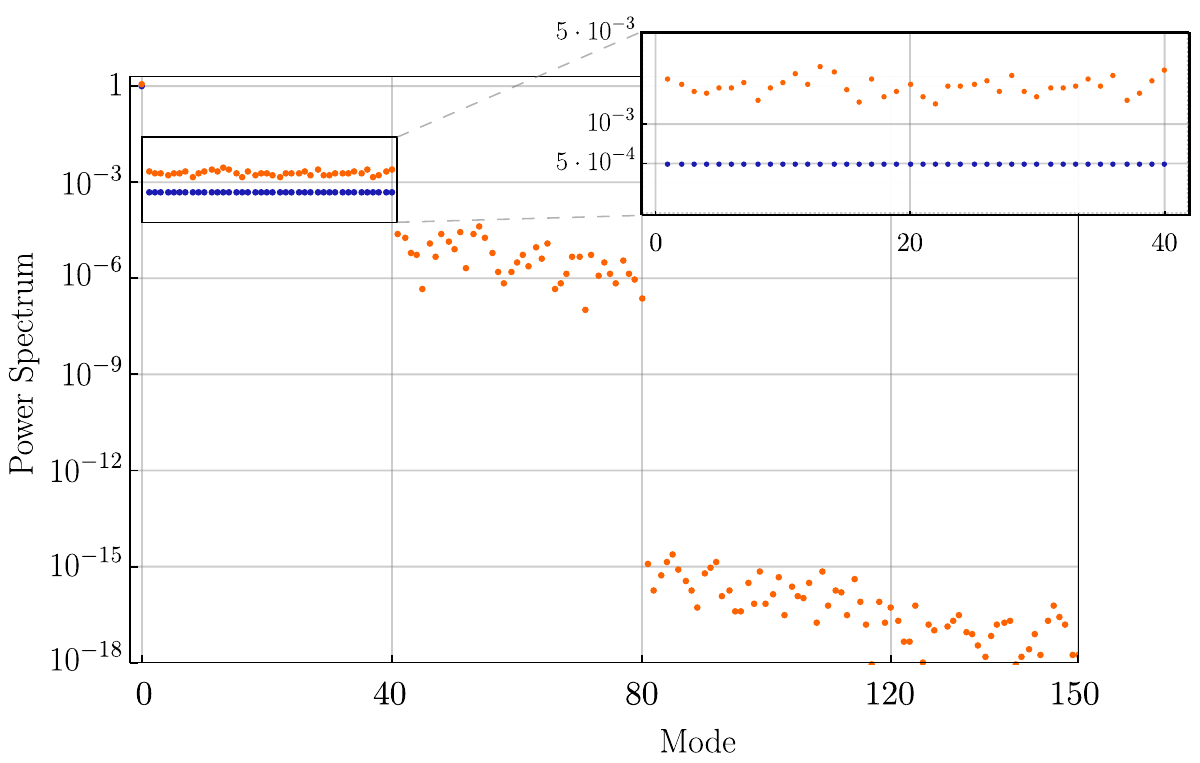}
    \caption{Power spectrum: square of the absolute value of the Fourier transform of $\mu(x)$ in blue and $\rho(x)$ in orange. 
    The transform is normalized so that the zeroth mode of $\mu(x)$ equals one. Unexcited modes of $\mu(x)$, i.e. those above $40$, lie below $10^{-30}$. }
    \label{fig:SpectrumMuRoAdS4}
\end{figure}

We begin by characterizing the inhomogeneous geometry corresponding to a typical realization of a disordered chemical potential.
In the left panel of Fig.~\ref{fig:dismurhoAdS4}, we show a realization of the noisy chemical potential~\eqref{eq:dismu} with $T/\mu_0= 0.06$ and disorder strength $V/\sqrt{\mu_0} =2.5$.
The corresponding charge density is plotted in the right panel.
The amplitude of the charge density oscillations is clearly amplified relative to that of the chemical potential.
This feature is confirmed by a spectral analysis in Fig.~\ref{fig:SpectrumMuRoAdS4}. There we show the power spectrum of both the chemical potential and the charge density 
from Fig.~\ref{fig:dismurhoAdS4}. As expected, the power spectrum of $\mu(x)$ is flat and, for this realization with $N=40$, contains 40 nonzero modes.
For the response $\rho(x)$,
the first 40 modes are amplified with respect to their counterparts in $\mu(x)$, higher modes up to $N=80$ are suppressed by at least two orders of magnitude,
and even higher modes are clearly negligible.
This amplification of noise in the response agrees with the findings of~\cite{Arean:2013mta,Hartnoll:2014cua}.

\begin{figure}
    \centering
 \begin{subfigure}[b]{0.49\linewidth}
            \centering
            \includegraphics[width=0.95\linewidth]{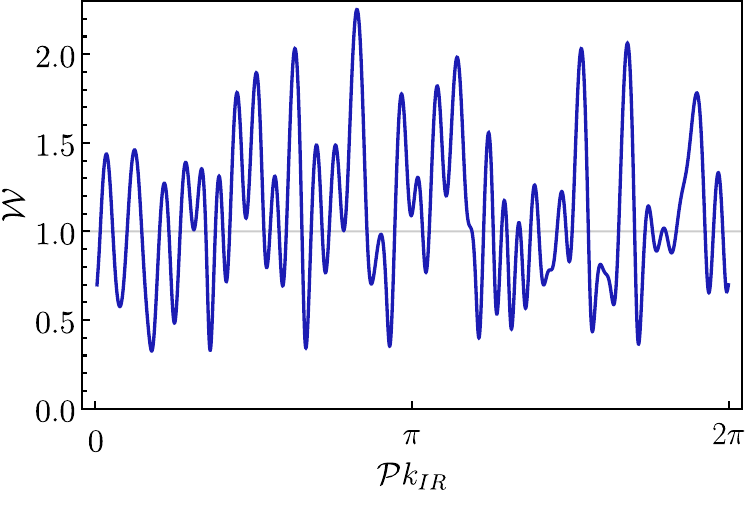}

        \end{subfigure}
        \begin{subfigure}[b]{0.49\linewidth}
            \centering
            \includegraphics[width=\linewidth]{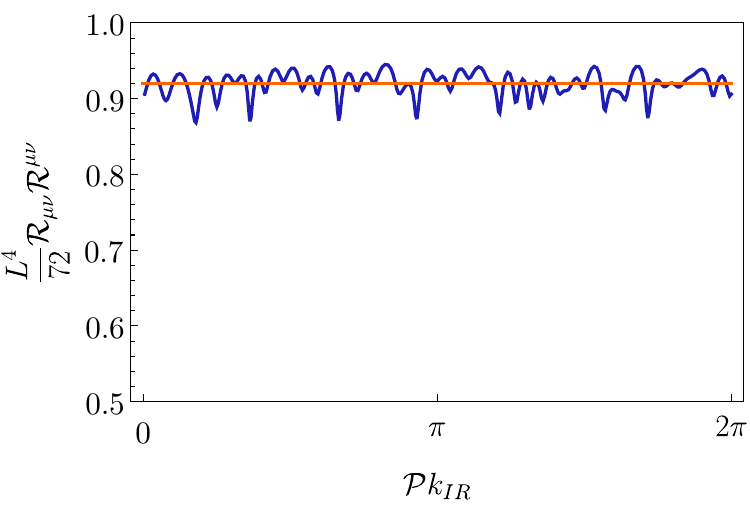}
        \end{subfigure}
    \caption{Disordered geometry near the horizon. Left: Norm of the Killing vector along the $y$ direction evaluated at the horizon and normalized such that $\lim_{z \to 0 } \left| \frac{z}{L}\partial_y\right|^2= 1 $, so in our ansatz ${\cal W}(z=1,x)=h_4(1,x)$. Right: Induced Ricci tensor squared at the horizon, normalized by the (clean) extremal value (shown in orange is the clean value for the same temperature). Both quantities are plotted as functions of $\mathcal{P}\,k_{\rm IR}$, the proper distance along the $x$ coordinate on the horizon.}
    \label{fig:DisorderGeometry}
\end{figure}

Next, we study the inhomogeneity of the geometry, with particular emphasis on the near-horizon region.
A useful quantity to illustrate this concept is the norm of the
Killing vector along the homogeneous $y$ direction, namely
\begin{equation}
\label{eq:calW}
{\cal W}= \left|{z\over L}\,\partial_y\right|^2\,,
\end{equation}
where we have normalized by the factor $z/L$ such that ${\cal W}$ approaches 1 at the boundary. We plot ${\cal W}$ at the horizon in the left panel of Fig.~\ref{fig:DisorderGeometry}, and in the right panel we show the induced Ricci tensor squared at the horizon. Both quantities show disordered oscillations around their values in the clean case. However, the inhomogeneities are much more pronounced in the Killing vector $\partial_y$: ${\cal W}$ fluctuates by about 95\% around its mean, whereas the norm of the Ricci tensor varies by
only about 5\%. This indicates that disorder introduces strong directional modulation without significantly altering the overall curvature scale. 
We return to this point below when studying the low temperature limit of these geometries.


\subsection{Low temperature solutions}
\label{ssec:ads4lowt}

In this section, we focus on the low temperature solutions.\footnote{We shall stress that the extremal $T=0$ limit remains outside of the disordered regime~\eqref{eq:discutoffs}. Only in the $N\to\infty$ limit would we be able to reach $T=0$. Thus the question of the nature of the extremal disordered horizon cannot be addressed in this analysis.}
In the clean case, the 
near-horizon geometry of the near-extremal AdS$_4$-RN black hole becomes
AdS$_2\times \mathbb{R}^2$.
We intend to analyze if our disordered solutions approach the AdS$_2\times \mathbb{R}^2$ in the low temperature limit. For a given disorder distribution, namely once $k_0$ and $N$ are fixed, we will study the near-horizon geometry as $T/\mu_0$ is lowered. 
Recall that the IR cutoff~\eqref{eq:discutoffs} sets the minimum temperature accessible to our disordered solution.
We therefore study various geometric invariants in the near-horizon region
as functions of temperature and compare them to their clean counterparts.
All results in this section correspond to a disorder realization of~\eqref{eq:dismu} with $N=40$ and $k_0/\mu_0=$0.08.
We consider disorder strengths in the range $w\in [0.004, 0.02]$  corresponding to
$V/\sqrt{\mu_0} \in [0.25,1.25]$.
To characterize the horizon geometry, we compute the average and standard deviation of geometric invariants at the horizon.
We define the horizon average of a quantity $X$ 
in terms of the proper distance along the horizon as
\begin{equation}
\label{eq:horizonav}
    \int_\mathcal{H} X = \frac{1}{\int dx\,dy\,\sqrt{\gamma_\mathcal{H}} } \int dx\, dy\,  \sqrt{\gamma_\mathcal{H}}\, X\,.
\end{equation}
Here $\gamma_\mathcal{H}$ denotes the determinant of the induced metric on the horizon, which for our ansatz reduces to $\gamma_\mathcal{H}=h_3\,h_4$.
For the standard deviation we use its natural definition in terms of the average defined above, namely
\begin{equation}
\label{eq:horizonSD}
\text{SD}\left(X\right)_{\mathcal H}=
\sqrt{\int_\mathcal{H} X^2-\left(\int_\mathcal{H}X\right)^2}\,.
\end{equation}

\begin{figure}
    \centering
    \begin{subfigure}[b]{0.49\linewidth}
            \centering
            \includegraphics[width=\linewidth]{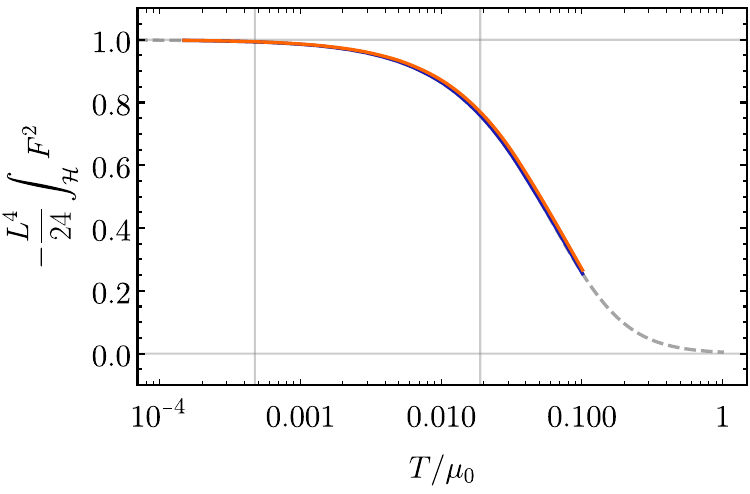}
            \captionsetup{justification=centering}

    \end{subfigure}
    \begin{subfigure}[b]{0.49\linewidth}
            \centering
            \includegraphics[width=\linewidth]{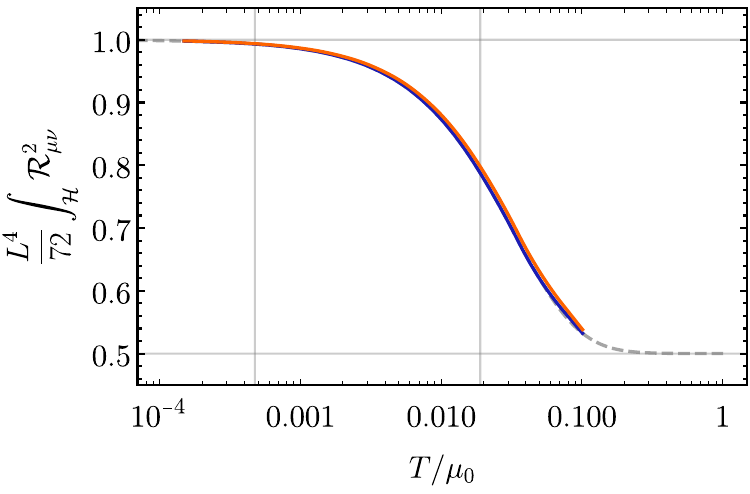}
        \end{subfigure}
    \caption{Spatial averages of $F^2$ (left) and $\mathcal{R}_{\mu\nu}\mathcal{R}^{\mu\nu}$ (right) at the horizon as functions of temperature. Blue and orange curves correspond to disorder strengths $V/\sqrt{\mu_0} = 0.25$ and $1.25$, respectively. Vertical gray lines mark the minimum and maximum temperatures accessible to our disorder distribution.The dashed gray line shows the result for the clean geometry.}
    \label{fig:LowT_AdS4_av1}
\end{figure}

In Fig.~\ref{fig:LowT_AdS4_av1} we plot the average values of $F^2$ and the curvature invariant
$\mathcal{R}_{\mu\nu}\mathcal{R}^{\mu\nu}$ at the horizon.
These averages coincide with those of the clean geometry
as shown by the fact that all curves overlap with that of AdS$_4$-RN.
Consequently, they scale with temperature as in the clean case, approaching
$F^2=24$ and $\mathcal{R}_{\mu\nu}\mathcal{R}^{\mu\nu}=72$, characteristic of the extremal black hole.
\begin{figure}
    \centering
 \begin{subfigure}[b]{0.49\linewidth}
            \centering
            \includegraphics[width=\linewidth]{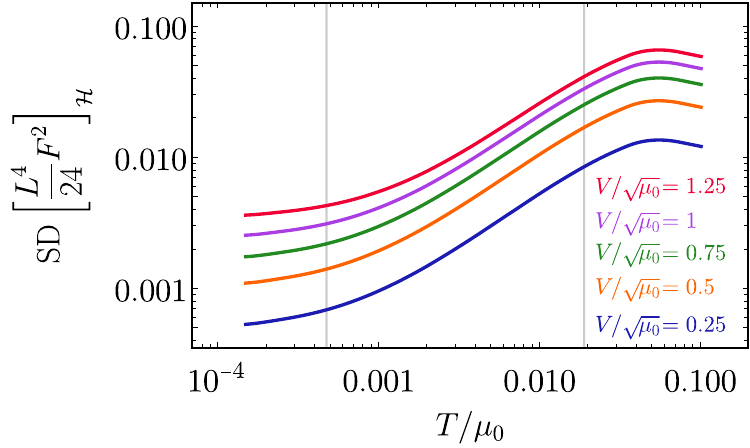}
            \captionsetup{justification=centering}

        \end{subfigure}
        \begin{subfigure}[b]{0.49\linewidth}
            \centering
            \includegraphics[width=\linewidth]{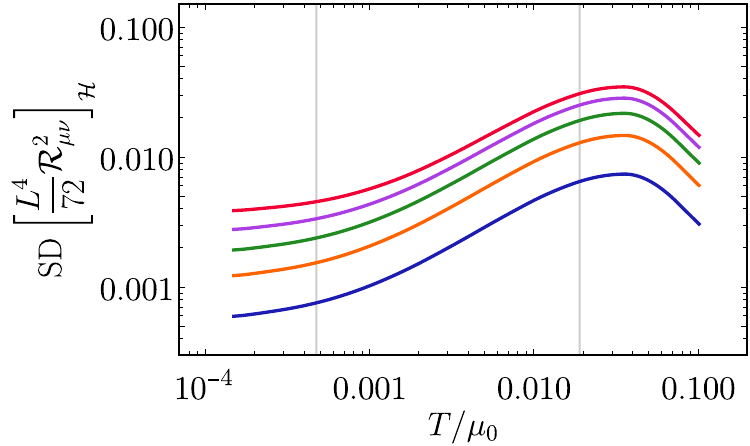}
        \end{subfigure}
    \caption{Standard deviations of $F^2$ (left) and $\mathcal{R}_{\mu\nu}\mathcal{R}^{\mu\nu}$ (right) for various disorder strengths $V/\sqrt{\mu_0}$, as indicated in the left panel.}
    \label{fig:LowT_AdS4_Stds_1}
\end{figure}
The  standard deviations of these observables are shown in Fig.~\ref{fig:LowT_AdS4_Stds_1}.
Although they decrease with temperature, the decline is slower than a power law, 
suggesting a residual value at vanishing temperature.
In Appendix~\ref{app:modads4} we study modulated geometries, namely solutions
where the chemical potential consists of a single cosine instead of the sum~\eqref{eq:dismu}. There, in agreement with the findings of~\cite{Hartnoll_2014flopy}, we confirm that for a range of values of $k_0/\mu_0$, the extremal horizon is modulated and thus features a nonzero value for the standard deviations in Fig.~\ref{fig:LowT_AdS4_Stds_1}.

\begin{figure}
    \centering
    \begin{subfigure}[b]{0.47\linewidth}
            \centering
            \includegraphics[width=\linewidth]{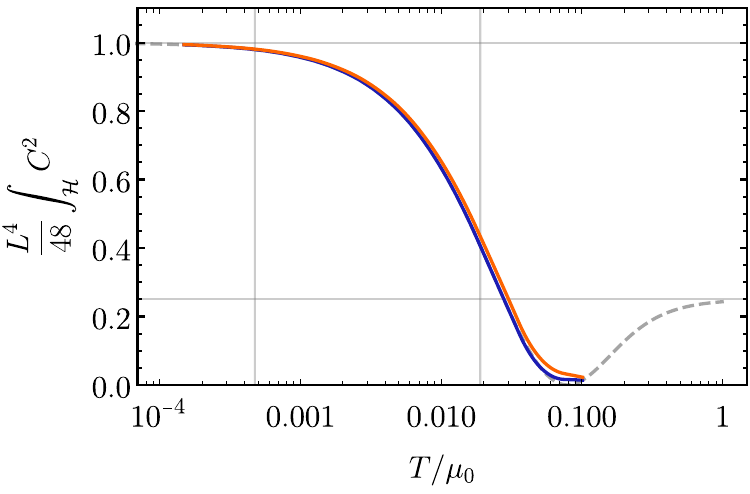}
            \captionsetup{justification=centering}

    \end{subfigure}
    \begin{subfigure}[t]{0.51\linewidth}
        \centering
        \includegraphics[width=\linewidth]{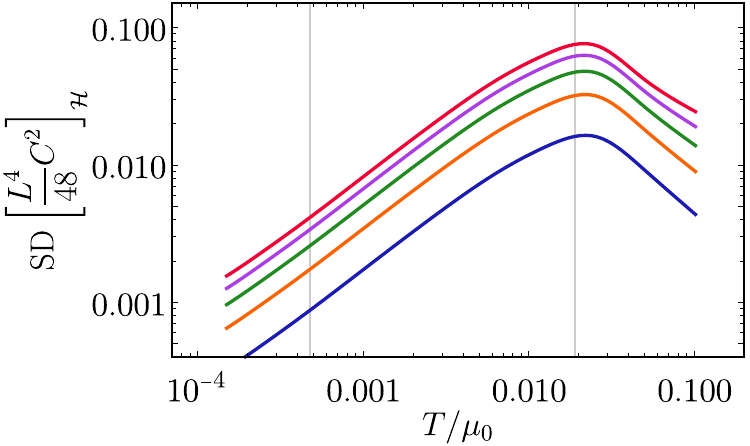}
    \end{subfigure}
    \caption{Left: Spatial average of the Weyl tensor $C$ squared at the horizon for the smallest (blue) and largest (red) noise strengths. The gray dashed line indicates the value for the clean geometry and the gray solid horizontal lines show the UV and IR values of $C^2$. 
    Right: Standard deviation of $C^2$ at the horizon for various disorder strengths. (Color code as in Fig.~\ref{fig:LowT_AdS4_Stds_1}.)}
    \label{fig:Weyl}
\end{figure}

Another interesting geometric observable is the Weyl tensor $C$.
It is independent of the Ricci tensor, depends on components of the Riemann tensor unconstrained by Einstein's equations, and
is invariant under conformal rescalings.
In Fig.~\ref{fig:Weyl} we plot the average and standard deviation of the Weyl tensor squared.
As with $F^2$ and $\mathcal{R}_{\mu\nu}\mathcal{R}^{\mu\nu}$
in Fig.~\ref{fig:LowT_AdS4_av1},
the average value of $C^2$ behaves as in the clean case.
The standard deviation, however, vanishes as a power law toward low temperature for all noise strengths.\footnote{Since the standard deviation of 
$C$ vanishes toward the IR and thus $C$ takes the AdS$_2$ value, one cannot exclude the possibility that our inhomogeneous IR geometry is conformal to AdS$_2 \times \mathbb{R}^2$.
Indeed, an $x$-dependent conformal transformation of the clean geometry would render $\mathcal{R}_{\mu\nu}\mathcal{R}^{\mu\nu}$ inhomogeneous and leave $C$ unchanged. }

\begin{figure}
    \centering
 \begin{subfigure}[b]{0.49\linewidth}
            \centering
            \includegraphics[width=\linewidth]{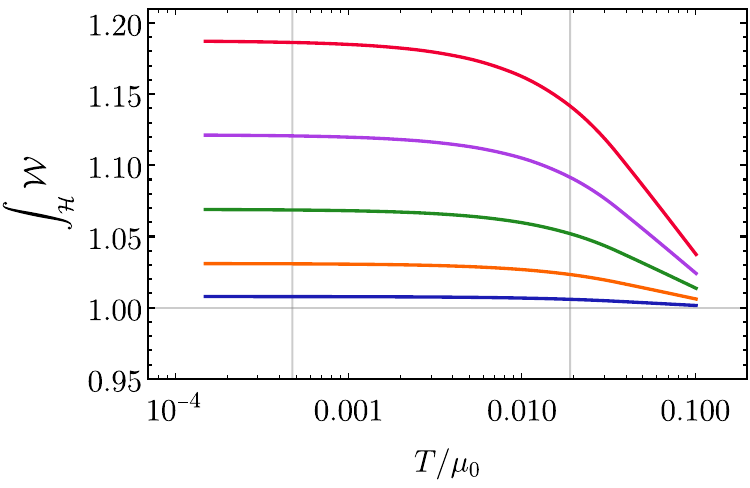}
            \captionsetup{justification=centering}

        \end{subfigure}
        \begin{subfigure}[b]{0.49\linewidth}
            \centering
            \includegraphics[width=0.95\linewidth]{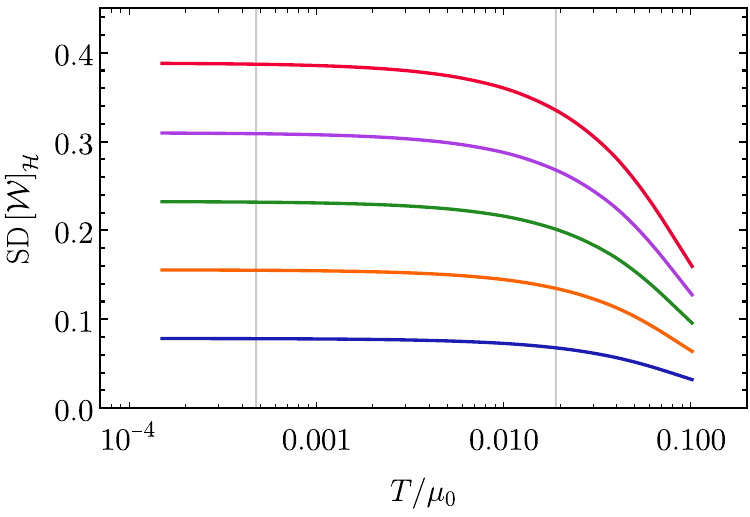}
        \end{subfigure}
    \caption{Average (left) and standard deviation (right) of ${\cal W}$ as functions of temperature for several disorder strengths, with the same color code as in Fig.~\ref{fig:LowT_AdS4_Stds_1}.}
    \label{fig:LowT_AdS4_curlyw}
\end{figure}

The last geometric invariant we examine is ${\cal W}$, the norm of the Killing vector along the remaining homogeneous spatial direction, defined in~\eqref{eq:calW}.
For our ansatz~\eqref{eq:ads4ansatz}, at the horizon this quantity reduces to
${\cal W}=h_4(z=1,x)$.
We study both the average and standard deviation of ${\cal W}$, with particular attention to the latter.
If the standard deviation approaches a nonzero value in the low temperature limit, 
it indicates that the disordered geometries do not flow to the clean fixed point.
In Fig.~\ref{fig:LowT_AdS4_curlyw} we plot the average of~${\cal W}$ (left panel) and its standard deviation (right panel) for different values of the disorder strength.
Crucially, we find that the standard deviation of ${\cal W}$ stabilizes at a nonzero value as temperature decreases.
As we will see in the next section, the persistence of a disordered horizon in the low temperature limit has direct implications for the scaling of the electrical resistivity.

\begin{figure}
    \centering
 \begin{subfigure}[b]{0.49\linewidth}
            \centering
            \includegraphics[width=\linewidth]{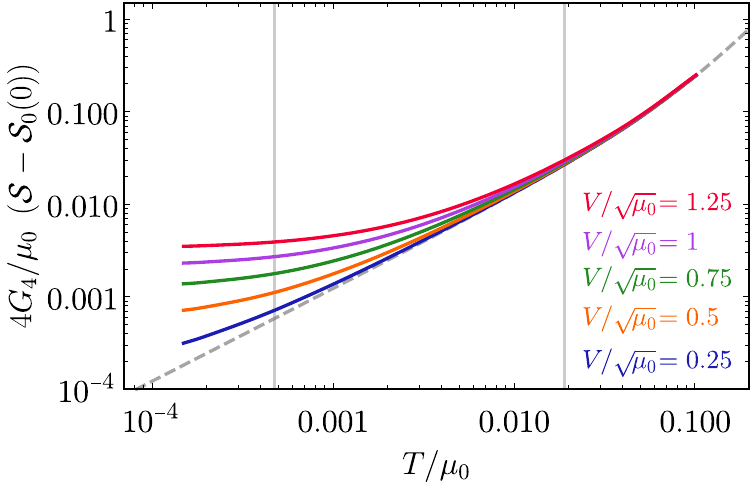}
            \captionsetup{justification=centering}

        \end{subfigure}
        \begin{subfigure}[b]{0.49\linewidth}
            \centering
            \includegraphics[width=0.975\linewidth]{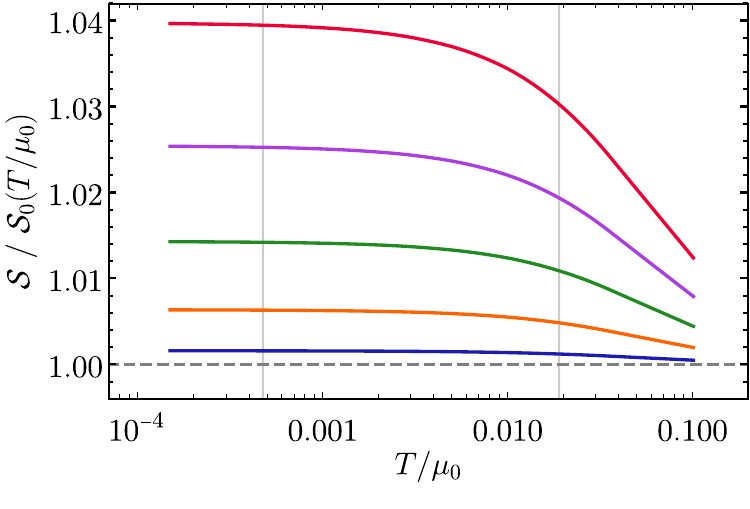}
        \end{subfigure}
    \caption{Entropy as a function of temperature for different disorder strengths. In the left panel, $4 G_4 \mathcal{S}_0 (0)=\frac{\mu_0^2}{12}$, corresponding to the clean extremal horizon, has been subtracted from all curves. In the right panel, we display the ratio between the entropy of the disordered and clean geometry.
   In both plots the gray dashed line corresponds to the clean solution AdS$_4$-RN. }
    \label{fig:LowT_AdS4_entropy}
\end{figure}

We conclude this subsection by examining the low temperature behavior of the entropy of our disordered configurations.
The temperature dependence of the entropy provides information about the existence and nature of disordered fixed points~\cite{Hartnoll:2015faa}.
The entropy density is given by the area of the horizon and, in terms of our ansatz~\eqref{eq:ads4ansatz}, can be written as
\begin{equation}
 \mathcal{S} =  \frac{\mathcal{S}_0}{L_x} \int dx \sqrt{h_3(1, x)\,h_4(1, x)} \ , \\
\end{equation}
where
\begin{equation}
    \frac{4 G_4}{\mu_0^2} \mathcal{S}_0    =  \left( \frac{2 \pi}{3} \frac{T}{\mu_0} + \sqrt{\left(\frac{2 \pi}{3} \frac{T}{\mu_0}\right)^2 + \frac{1}{12}} \right)^2 
\end{equation}
is the entropy of the clean case, \textit{i.e.} the AdS$_4$-Reissner-Nordstr\"om black brane. 
We plot the entropy in Fig.~\ref{fig:LowT_AdS4_entropy}.
In the left panel, we show the temperature dependence of the entropy 
after subtracting its value for the (clean) extremal AdS$_4$-RN geometry.
We observe that, as in the modulated case~\cite{Donos:2014yya}, in the low temperature limit the entropy asymptotes to a value larger than the clean extremal one. In the right panel, we present the ratio of the disordered and clean case entropies as a function of temperature.
This ratio flattens out at low temperatures, indicating that the entropy scales linearly with temperature for all disorder strengths.
Consistent with our results for the averaged geometry,
the entropy behaves in the low $T$ limit as in the clean system. It scales linearly and tends to a constant value that is enhanced as disorder is increased. This enhancement can be understood as a disorder-dependent length renormalization in the IR.

\subsubsection{IR deformation}
\label{sssec:AdS4IRDef}

\begin{figure}
    \centering
    \includegraphics[width=0.6\textwidth]{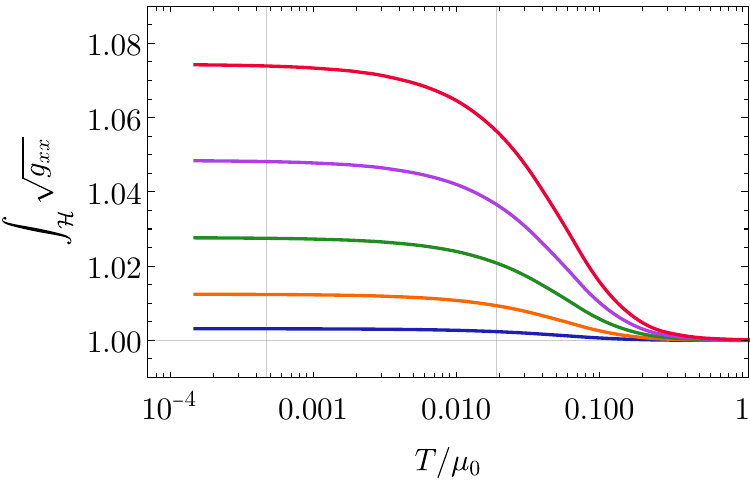} 
    \caption{$\sqrt{g_{xx}}$ integrated along the horizon for 
    different values of the disorder strength (according to the same color code as in Fig.~\ref{fig:LowT_AdS4_entropy}). Grey vertical lines mark the boundaries of the disorder regime. This integral measures the wavelength renormalization of the system.}
    \label{fig:Wave_Renorm}
\end{figure}

Figures~\ref{fig:LowT_AdS4_av1} and \ref{fig:Weyl} show that in the low $T$ limit, the near-horizon geometry averages to the clean critical point AdS$_2\times\mathbb{R}^2$.
This suggests that in the IR disorder can be treated as
a small deformation of the AdS$_2$ throat, effectively renormalizing 
the transverse directions, as illustrated by Fig~\ref{fig:LowT_AdS4_curlyw}.
To test the  consistency of this picture,
we study the conditions under which a small inhomogeneous perturbation of the gauge field $\delta A_t(r,x)=a_t(r)\,\cos(k\,x)$, where $r$ is the radial coordinate of the AdS$_2\times\mathbb{R}^2$ geometry,  can deform the throat.
Following~\cite{Hartnoll:2012rj}\footnote{Naturally, as in ~\cite{Hartnoll:2012rj} the perturbation of $A_t$ couples to those of the metric and the component of the gauge field along the disordered direction, $A_x$.}
one finds that towards the boundary of the AdS$_2$ throat (i.e. $r\to\infty$)
\begin{equation}
a_t(r) \propto r^{\Delta -1 } \quad \text{  with  } \quad \Delta = 1 + \frac{1}{2} \hat{k}^4 +\mathcal{O}(\hat{k}^6 )\,,
\label{eq:IRdimAdS4}
\end{equation}
where $\hat{k}$ is the perturbation wavevector in AdS$_2$ units. In our inhomogeneous geometries we can relate the wavelength of the perturbation in the IR ($\hat{\lambda}$) to the wavelength in the UV via 
\begin{equation}
    \hat{\lambda}= \sqrt{\left. g_{xx}\right|_\mathcal{H}} \ \lambda \ .
\end{equation}
The renormalization of lengths in the IR of the disorder theory would result from averaging
$\sqrt{\left. g_{xx}\right|_\mathcal{H}}$ over disorder realizations.
We consider the horizon average as a proxy and plot it in Fig.~\ref{fig:Wave_Renorm}.
For the strongest disorder, the wavenumber is renormalized by an order-one factor of $1/1.08\approx 0.92$. Moreover, we see that this value is stable at low temperature within the disorder regime.
This result implies that the wavenumber of a UV perturbation is not significantly renormalized, nor a measurable temperature dependence is introduced along the flow towards the IR.
Therefore, in the limit $\hat{k}\to0$, a modulated chemical potential would act as a marginal deformation of the IR AdS$_2\times\mathbb{R}^2$ fixed point.
We will see in Section~\ref{sssec:AdS3IRDef} that this is not the case for AdS$_3$ theories.


\subsection{DC conductivities}
We now study  DC thermoelectric transport in the AdS$_4$ disordered geometries. As discussed in Section~\ref{ssec:DCtransport}, the DC thermoelectric conductivities can be expressed in terms of horizon data.

\begin{figure}[htb]
    \centering
         \begin{subfigure}[t]{0.45\linewidth}
            \centering
            \includegraphics[width=\linewidth]{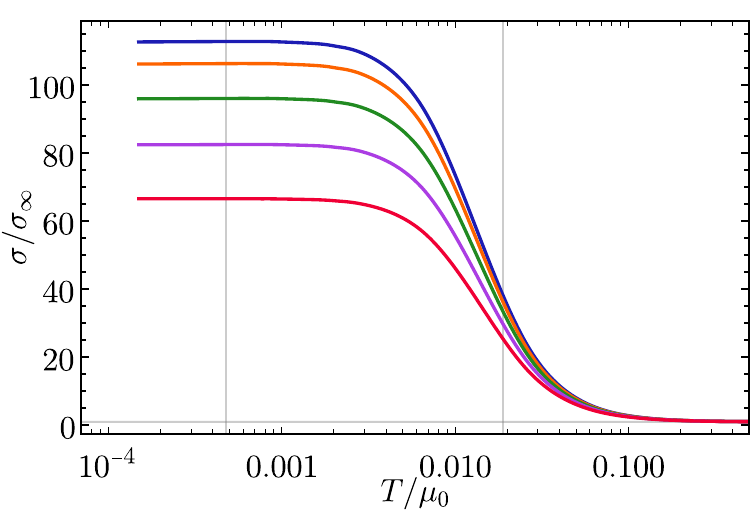}
        \end{subfigure}
        \begin{subfigure}[t]{0.51\linewidth}
            \centering
            \includegraphics[width=\linewidth]{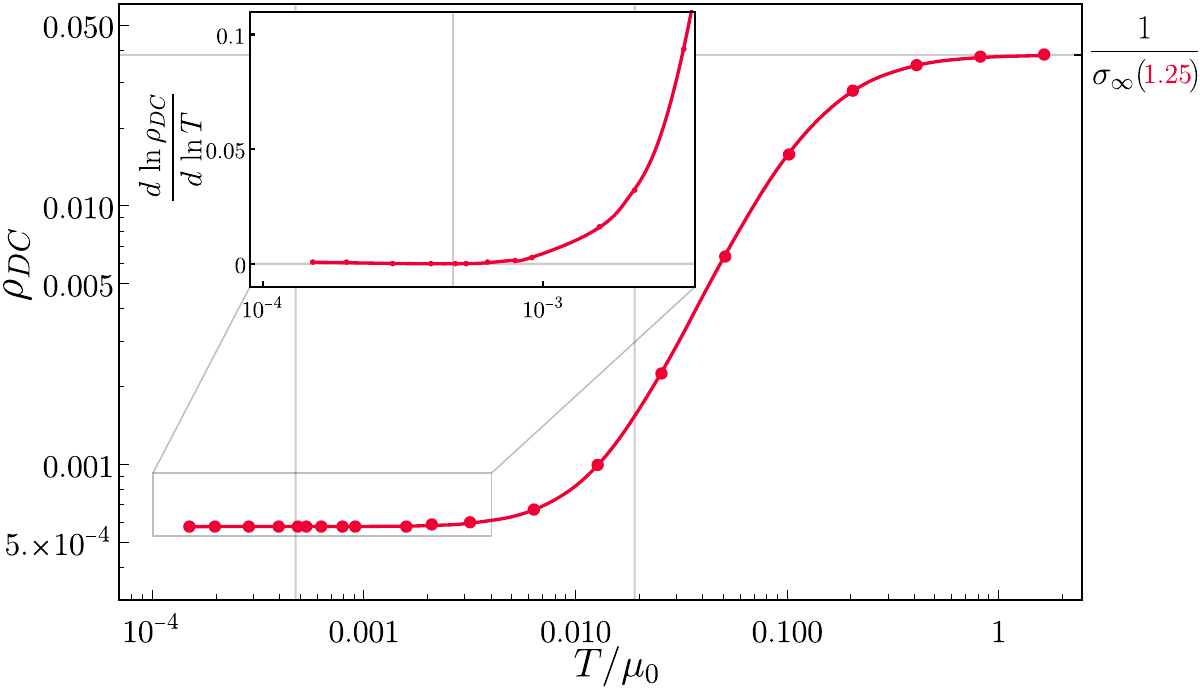}
            \captionsetup{justification=centering}

        \end{subfigure}
    \caption{Left: DC electrical conductivity $\sigma$ normalized to its value at infinite temperature~\eqref{eq:cond_inf} (with color code as in Fig.~\ref{fig:LowT_AdS4_entropy}). Vertical gray lines mark the boundaries of the disordered regime. Right: Resistivity $\rho_{DC}=1/\sigma$ as a function of temperature for the largest disorder strength, $V/\sqrt{\mu_0}=1.25$. The gray horizontal line is the infinite-temperature value of $\sigma$ for this disorder strength. The inset shows the logarithmic derivative of $\log\rho_{DC}$; at low $T$ it is consistent with a residual resistivity.}
\label{fig:DCAdS4}
\end{figure}

In the left panel of Fig.~\ref{fig:DCAdS4} we plot the electrical DC conductivity $\sigma$ of the AdS$_4$ black holes. 
As expected, at fixed temperature the conductivity is lower for larger disorder.
Remarkably, at low temperatures $\sigma$ approaches a constant,
indicating that disorder induces a residual resistivity. 
This can be better understood from the explicit form of the conductivity
in~\eqref{eq:sigmaDCAdS4}:
$\sigma$ would diverge if our geometry flowed to the clean extremal AdS$_4$-RN solution,
since as $V$ decreases and the near-horizon geometry becomes homogeneous,
$A_t$ and the metric become homogeneous, thus $\langle\rho^2\rangle\to\langle\rho\rangle^2$ and $\left<\Upsilon^2\right>\to0$. The dual theory would then behave as a clean metal at low temperature. However, as we discuss below, the disordered nature of the horizon is responsible for the residual resistivity we observe in the right panel of Fig.~\ref{fig:DCAdS4}. There we plot $\rho_{DC} = 1/\sigma$ 
for the largest disorder in our dataset. We find that at low temperatures within the disordered regime the resistivity is approximately
constant\footnote{{In~\cite{Hartnoll:2012rj} it was predicted that disorder leads to a low $T$ scaling of the form $\rho_{DC}\sim1/\log T$. However this scaling would hold at extremely low temperatures $T\sim e^{-1/k_{\rm IR}}$ which lie outside the disorder
regime \eqref{eq:discutoffs}.}}
within our numerical precision.

\begin{figure}[htb]
    \begin{subfigure}[b]{0.49\linewidth}
        \centering
        \includegraphics[width=\linewidth]{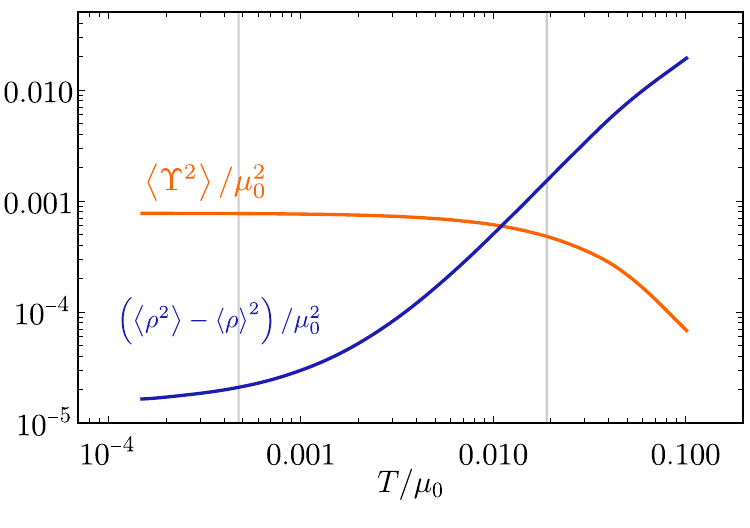}
    \end{subfigure}
    \begin{subfigure}[b]{0.51\linewidth}
        \centering
        \includegraphics[width=\linewidth]{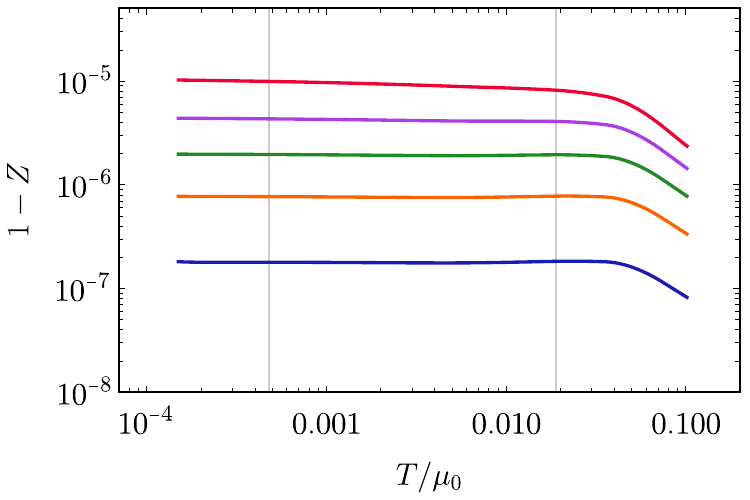}
    \end{subfigure}
    \caption{Contributions to the transport coefficients in \eqref{eq:sigmaDCAdS4} as functions of temperature. Left: The two terms in the denominator.
     At high temperatures the inhomogeneity of the electric field dominates, while at low temperatures 
    the inhomogeneous geometry is the leading contribution.
    Right: The overall normalizing factor in the conductivity,
    which remains below one, consistent with \cite{Grozdanov:2015djs}.} 
   \label{fig:contribcond}
\end{figure}

In the left panel of Fig.~\ref{fig:contribcond} we analyze the two contributions to the residual resistivity given by the denominator of~\eqref{eq:sigmaDCAdS4}.
First, $\langle\rho^2\rangle-\langle\rho\rangle^2$ is a measure of the standard deviation of the electric field at the
horizon.\footnote{It is not exactly SD$(\rho)$ because the average defining the standard deviation~\eqref{eq:horizonSD} differs from $\langle\rangle$ in \eqref{eq:ads4xaverage}
by a factor of $h_4$ in the measure. Using Hölder's inequality, one can check that the two averages differ by at most a factor of $\int_\mathcal{H}1/h_4 < 1$.}
Second, $\left<\Upsilon^2\right>$, defined in
\eqref{eq:rho_upsil_def}, encodes the inhomogeneity of the horizon
along the spatial direction $y$ orthogonal to the disorder, 
and is therefore related to the norm of $\partial_y$ studied in Fig.~\ref{fig:LowT_AdS4_curlyw}.
Our data shows that the electric field contribution decreases towards low temperature, 
slower than a power law, consistent with Fig.~\ref{fig:LowT_AdS4_Stds_1}. 
Instead, $\left<\Upsilon^2\right>$ initially grows as temperature is lowered, flattening out in the low temperature regime.
This mimics the behavior observed for the standard deviation of ${\cal W}$
in Fig.~\ref{fig:LowT_AdS4_curlyw}.
Finally, in the right panel of Fig.~\ref{fig:contribcond} we plot the overall factor $Z$ appearing in the expression of the DC conductivity~\eqref{eq:sigmaDCAdS4}. For $\sigma$ to satisfy the lower bound proved in~\cite{Grozdanov:2015qia}, even in the hypothetical case where disorder is so strong in the IR that the denominator of the second contribution to $\sigma$ vanishes, it must be that $Z>1$.
As the plot shows, $Z$ is above one for all temperatures and disorder strengths. For larger disorder it saturates to higher values in the low temperature regime.

The denominator of the DC conductivities~\eqref{eq:sigmaDCAdS4} admits an interpretation in terms of massive gravity.
As shown in~\cite{Blake:2013owa}, inhomogeneities in the dual field theory break translational invariance in the bulk, giving the graviton an effective mass.
Comparing \eqref{eq:sigmaDCAdS4} with the conductivity in massive gravity models \cite{Blake:2013bqa}, 
we infer the would-be effective mass for the graviton at the horizon
\begin{equation}
    M_g^2  \approx {\langle\rho^2\rangle - \langle\rho\rangle^2+\langle\Upsilon^2\rangle} \, .
\end{equation}
The massive graviton is responsible for dissipation and determines the relaxation rate in the dual field theory.

\begin{figure}
    \centering
 \begin{subfigure}[b]{0.49\linewidth}
            \centering
            \includegraphics[width=\linewidth]{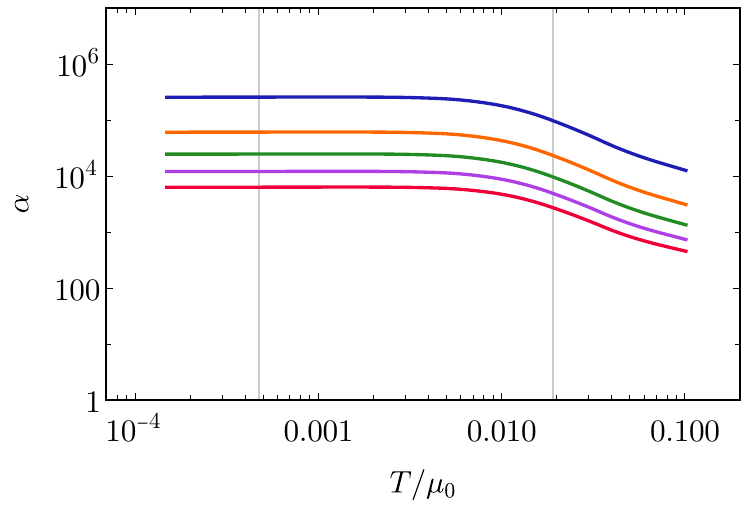}
            \captionsetup{justification=centering}

    \end{subfigure}
    \begin{subfigure}[b]{0.49\linewidth}
            \centering
            \includegraphics[width=\linewidth]{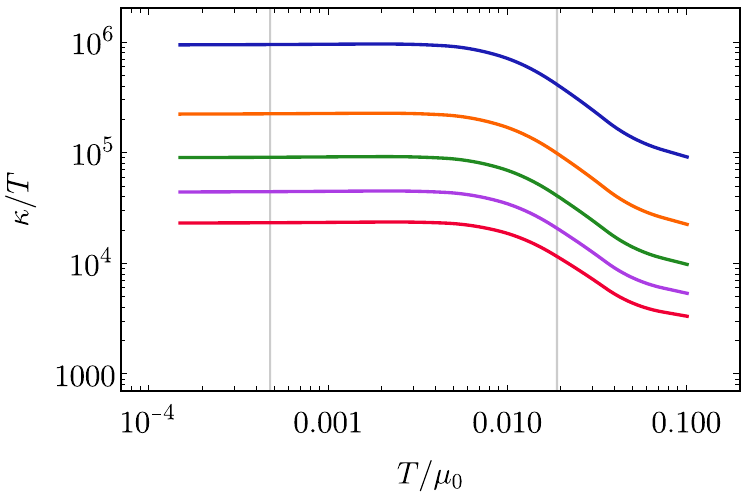}
        \end{subfigure}
    \caption{Left: Thermoelectric DC conductivity as a function of temperature for different noise strengths. Right: Thermal DC conductivity normalized by temperature for the same set of noise strengths.
    (Color code in both plots as in Fig.~\ref{fig:LowT_AdS4_entropy})}
    \label{fig:ThermoElectric}
\end{figure}

Finally, Figure~\ref{fig:ThermoElectric} shows the thermoelectric and thermal DC conductivities, $\alpha$ and $\kappa$, respectively.
At low temperatures $\alpha$ mirrors $\sigma$ and 
approaches a residual value that decreases with increasing disorder strength.
By contrast, $\kappa$ vanishes linearly as $T\to0$. 
Both behaviors follow from \eqref{eq:sigmaDCAdS4}:
$\alpha$ and $\kappa$ share the same denominator as $\sigma$, so the persistent inhomogeneity of the low $T$ horizon yields a finite value of $\alpha$ in the IR.
$\kappa$ carries an extra power of $T$ relative to $\alpha$ and $\sigma$,
which accounts for the linear vanishing of the thermal conductivity observed in the right panel of Fig.~\ref{fig:ThermoElectric}.

\subsection{Realization dependence}

\begin{figure}
    \centering
 \begin{subfigure}[b]{0.49\linewidth}
            \centering
            \includegraphics[width=\linewidth]{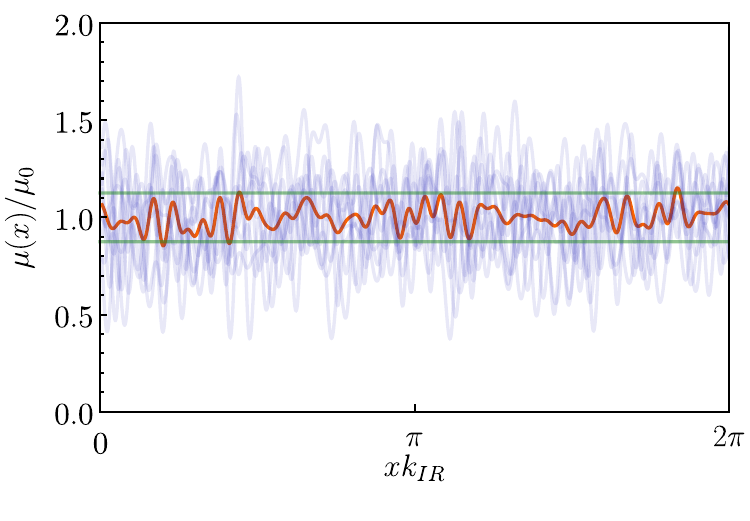}
            \captionsetup{justification=centering}

        \end{subfigure}
        \begin{subfigure}[b]{0.49\linewidth}
            \centering
            \includegraphics[width=\linewidth]{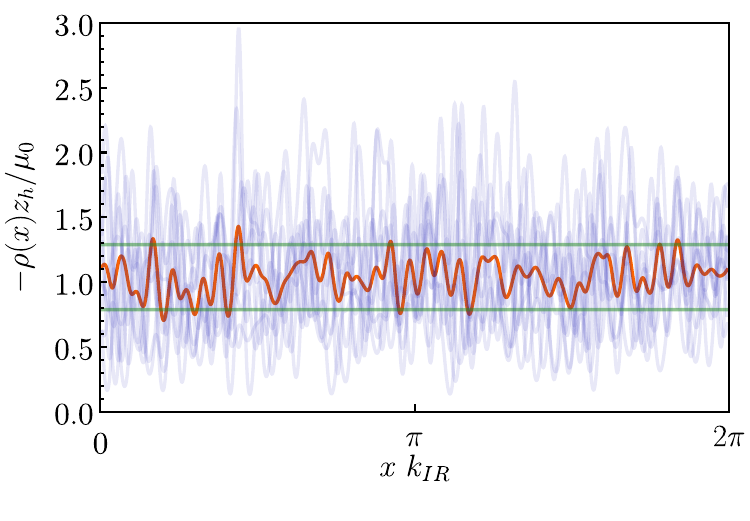}
        \end{subfigure}
   \caption{Left: Chemical potential $\mu(x)$. Right: Charge density $\rho(x)$. Light blue curves show $N_D=10$ realizations with identical parameters: $k_{UV}/\mu_0 = 0.08$, $N = 40$, $V/\sqrt{\mu_0} = 1.25$, and $T/\mu_0 = 0.0004$ (slightly below the lower edge of the disorder range). 
    The orange curve shows the disorder-averaged value at each point.
   These disorder-averaged profiles have the same mean as an individual realization,
    while the amplitude of their oscillations is suppressed by a factor of $1/\sqrt{N_D}$
    relative to the mean of the standard deviations.
    The green bands indicate the 95\% confidence interval around the mean, obtained as
    $1/\sqrt{N_D}$ times the mean over realizations of the standard deviation.
    Notice that indeed most of the orange lines fit within the green bands.}
   \label{fig:MultipleRealisations}
\end{figure}

In principle, a proper treatment of disorder requires many simulations, each with a different set of random phases $\{\delta_n\}$ of (\ref{eq:dismu}).
Physical observables should be averaged over these disorder realizations and depend only on properties of the disorder distribution (strength, correlations, $k_{UV}$), not on a particular choice of $\mu(x)$.
Unfortunately, numerical disorder averaging is costly.
If an observable is itself a sum of cosines with random phases, the average converges
as $1/\sqrt{N_D}$, where $N_D$ is the number of realizations.
Figure~\ref{fig:MultipleRealisations} illustrates this by averaging over ten realizations:
the standard deviation of the averaged signal is only a factor 
$0.3\approx 1/\sqrt{10}$ smaller than the average of the individual standard deviations, as expected.

\begin{figure}
    \centering
     \begin{subfigure}[b]{0.49\linewidth}
            \centering
            \includegraphics[width=\linewidth]{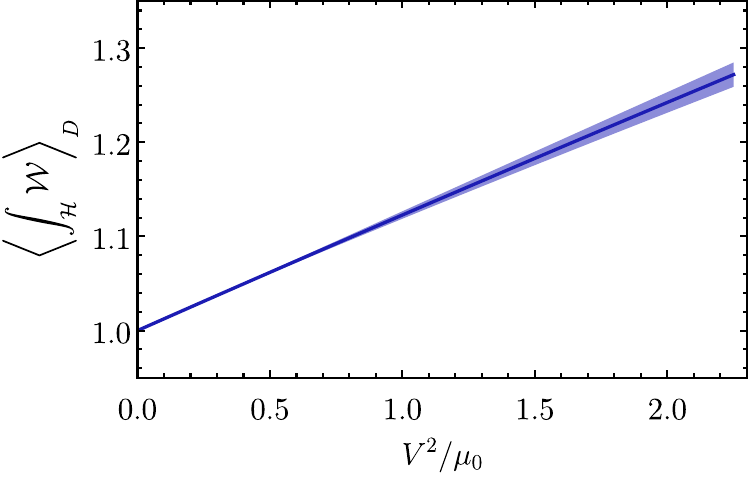}
            \captionsetup{justification=centering}

    \end{subfigure}
    \begin{subfigure}[b]{0.48\linewidth}
            \centering
            \includegraphics[width=\linewidth]{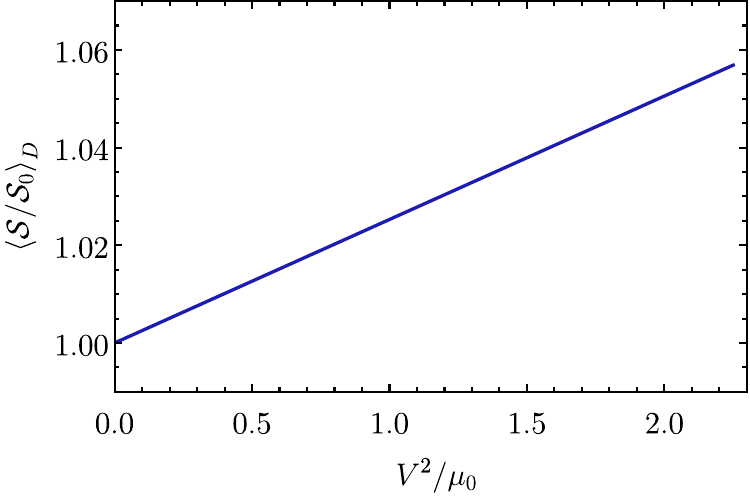}
    \end{subfigure}
    \caption{Disorder-averaged values of $\int_{\cal H}\mathcal{W}$ (left) and entropy density (right) as functions of the squared disorder strength.
    Results are shown for $N_D=10$ realizations with identical parameters:
   $k_{UV}/\mu_0 = 0.08$, $N = 40$, and $T/\mu_0 = 0.0004$
    (slightly below the lower edge of the disorder range).
   The line thickness indicates the standard deviation of the disorder ensemble.}
    \label{fig:DisorderAverages}
\end{figure}

To mitigate the issue of realization dependence we focus on
quantities integrated over $x$.
These ``self-averaging'' observables are insensitive to the random phases
$\{\delta_n\}$ (in the limit of infinitely many modes).
In Fig.~\ref{fig:DisorderAverages} we plot the mean and standard deviation
over ten disorder realizations of $\int_{\cal H}\mathcal{W}$ and the entropy ${\cal S}$ versus disorder strength.
The standard deviation over the ensemble of these self-averaging quantities is quite small, particularly for lower noise strengths,
indicating that a single realization closely approximates the disorder-averaged value.
Therefore, DC conductivities, entropy, and averaged geometric quantities that involve integration over $x$ do not change qualitatively upon disorder averaging.

\begin{figure}
    \centering
     \begin{subfigure}[b]{0.49\linewidth}
            \centering
            \includegraphics[width=\linewidth]{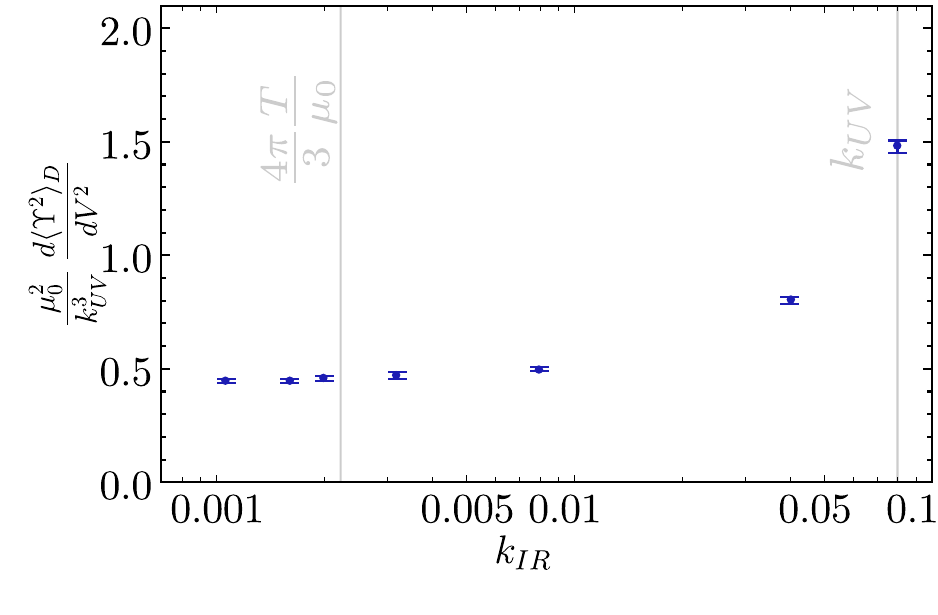}
            \captionsetup{justification=centering}
    \end{subfigure}
    \begin{subfigure}[b]{0.49\linewidth}
            \centering
            \includegraphics[width=\linewidth]{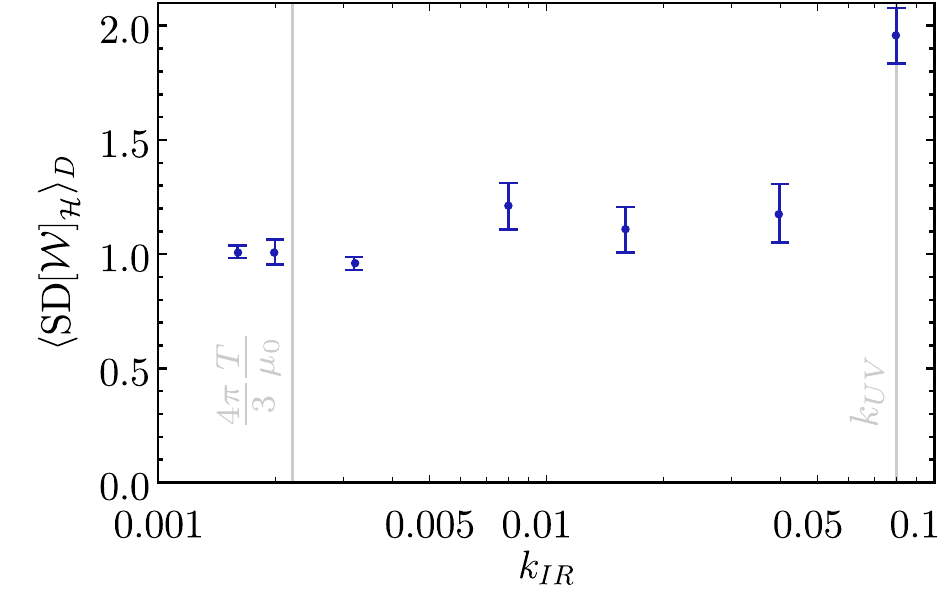}
    \end{subfigure}
    \centering
    \caption{Convergence of the derivative of the disorder-averaged $\left<\Upsilon^2\right>$ with respect to $V^2$ (left panel) and the standard deviation of $\mathcal{W}$ (right panel) as the number of modes is increased while holding temperature and $k_{UV}$ fixed (equivalently, as the IR cutoff $k_{IR}$ is taken to 0).
   The first quantity reflects the growth of the effective graviton mass with increasing disorder strength, whilst the second denotes the inhomogeneity of the horizon.}
  \label{fig:kir_convergence}
\end{figure}

One might also ask how our results change upon increasing the number of modes in the disordered source, or, equivalently, how observables depend on the IR cutoff $k_{IR}$.
In the left panel of Fig.~\ref{fig:kir_convergence} we show how the derivative of 
$\left< \langle\Upsilon^2\rangle \right>_D$
with respect to the disorder strength $V^2$ changes as we include more modes and $ k_{IR}$ decreases. Note that this derivative determines how the residual resistivity depends on the noise strength.
As the IR cutoff approaches the lower bound of the disordered regime, $\frac{4\pi}{3} \frac{T}{\mu_0}$, the derivative stabilizes.
In the right panel of Fig.~\ref{fig:kir_convergence} we repeat the analysis for the standard deviation of ${\mathcal W}$, finding that it also stabilizes as the number of modes becomes large enough.
We find similar results for a wide range of observables. 
Consequently, provided we remain inside the disordered regime, our solutions are
relatively insensitive to the IR cutoff and the qualitative results do not change as we
include more modes (equivalently, as $k_{IR} \to 0$).

Our results in the low temperature limit indicate 
that the IR of the averaged theory corresponds to a line of
critical points labeled by the noise strength.
Curvature invariants of the averaged geometry approach those of the clean quantum critical point, AdS$_2 \times \mathbb{R}^2$.
However, the averaged geometry is not itself a solution of Einstein gravity and is better interpreted as arising from an anisotropic massive-gravity–type effective theory.
The mass of the graviton in this averaged effective theory is the imprint of the inhomogeneous horizons featured in each individual realization and renders physical observables like transport coefficients finite.


\section{Disordered black holes in AdS$_3$}
\label{sec:ads3}

In this section, we present results for AdS-BTZ charged black holes in presence of disorder.
From the field theory perspective, we expect outcomes similar to those of the previous section, since disorder in AdS$_3$ violates the Harris criterion as in AdS$_4$. 
On the gravity side, however, we expect
significant differences, as gravity lacks propagating degrees of freedom in three dimensions.

\begin{figure}
    \centering
    \begin{subfigure}[b]{0.49\linewidth}
        \includegraphics[width=0.975\linewidth]{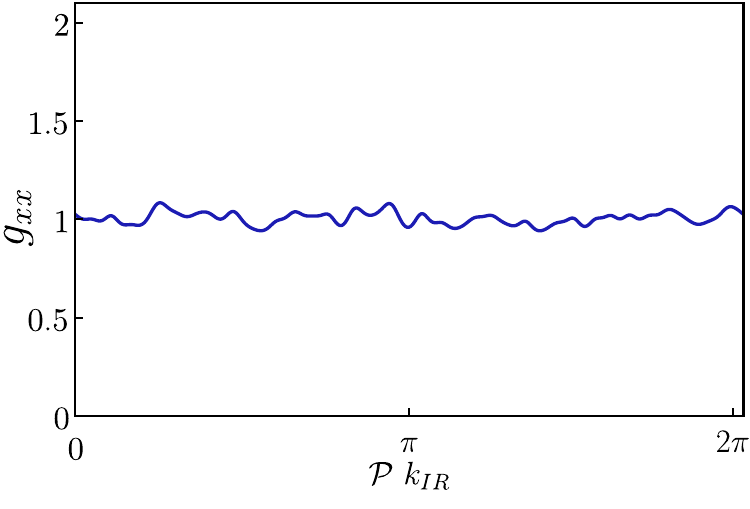}
    \end{subfigure}
    \begin{subfigure}[b]{0.49\linewidth}
        \includegraphics[width=\linewidth]{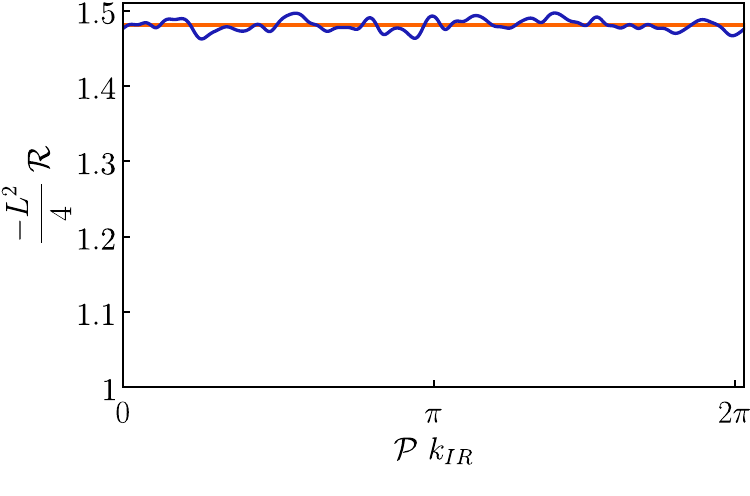}
    \end{subfigure}

    \begin{subfigure}[b]{0.49\linewidth}
        \includegraphics[width=\linewidth]{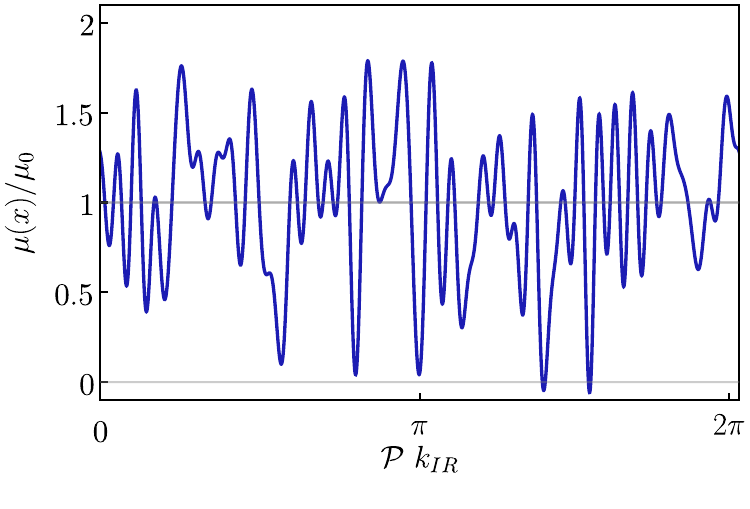}
    \end{subfigure}
    \caption{Inhomogeneous horizon of a BTZ black hole at high temperature, $T/\mu_0 \approx k_{UV} = 1$. 
    Top left: $g_{xx} $ component of the metric at the horizon. 
    Top right: Ricci scalar at the horizon (blue line), with the clean value at the same temperature shown in orange.
    Both plots are shown as functions of proper distance along the horizon for the same noise realization with $V/\sqrt{\mu_0}= 8$, displayed in the bottom panel.}
    \label{fig:ads3highT}
\end{figure}

We begin by presenting a typical realization of a noisy chemical potential in the bottom row of Fig.~\ref{fig:ads3highT}.
These data correspond to a high temperature configuration with $T/\mu_0=0.85$ and
$\mu(x)$ given by~\eqref{eq:dismu} with $k_{UV}=1$. 
As is evident from the plot, this corresponds to a strongly disordered source with oscillations of the same order as its mean.
The resulting inhomogeneous horizon geometry is shown in the upper row of 
Fig.~\ref{fig:ads3highT}.
Both the metric component $g_{xx}$ (left panel) and the Ricci scalar at the horizon (right panel) are noticeably less inhomogeneous than the source $\mu(x)$.
As we show below, this is a feature of disordered
charged BTZ black brane geometries, in which spatial inhomogeneities decay toward the IR.


\subsection{Low temperature solutions}

We now characterize the geometry of the low temperature solutions.
In the clean case these approach an AdS$_2 \times \mathbb{R}$ near-horizon geometry as $T\to0$.
With disorder, the evolution of the low temperature geometry differs markedly from that
of the AdS$_4$ black brane.
In fact, our findings run counter to the expectations implied by the Harris criterion: disorder does not affect the infrared geometry of charged AdS$_3$ black branes.

\begin{figure}
    \centering
    \includegraphics[width=0.6\linewidth]{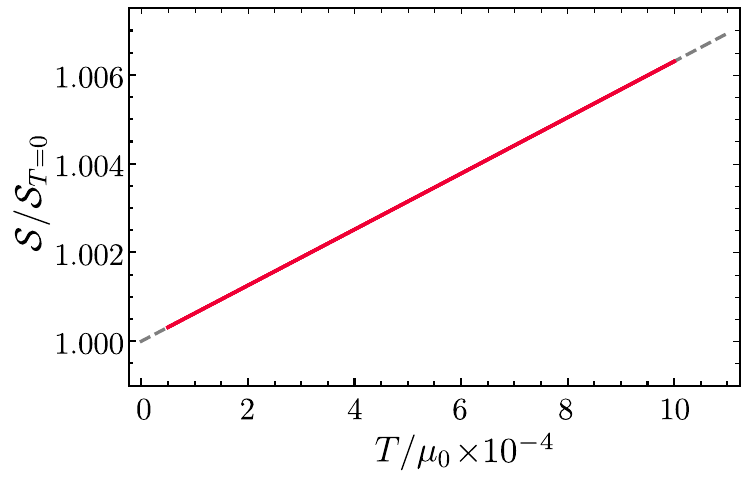}
    \caption{Entropy of the disordered AdS$_3$ BTZ black hole normalized by its value at zero temperature for disorder strengths
    $V/\sqrt{\mu_0}$ in the range $[0.2,1]$ (see Fig.~\ref{fig:AdS3_Ricci_Mean_Std}).
    Entropy is independent of disorder strength: all curves overlap and follow the clean case (gray dashed line).}
    \label{fig:AdS3Entropy}
\end{figure}

\begin{figure}
    \centering
    \begin{subfigure}[b]{0.49\linewidth}
        \centering
        \includegraphics[width=\linewidth]{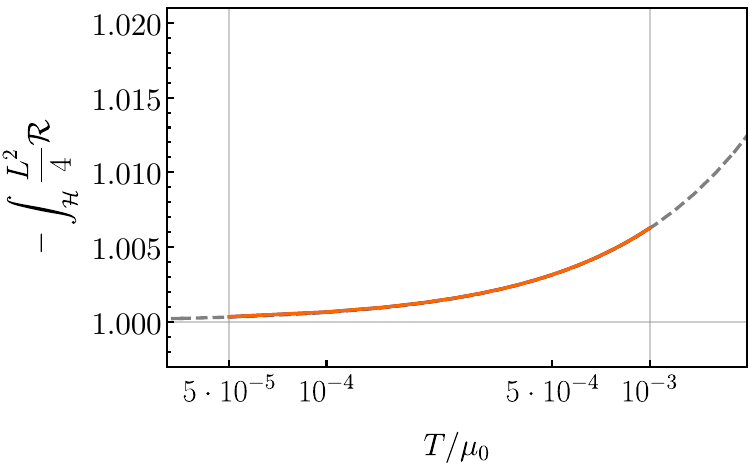}
    \end{subfigure}
    \begin{subfigure}[b]{0.49\linewidth}
        \centering
        \includegraphics[width=\linewidth]{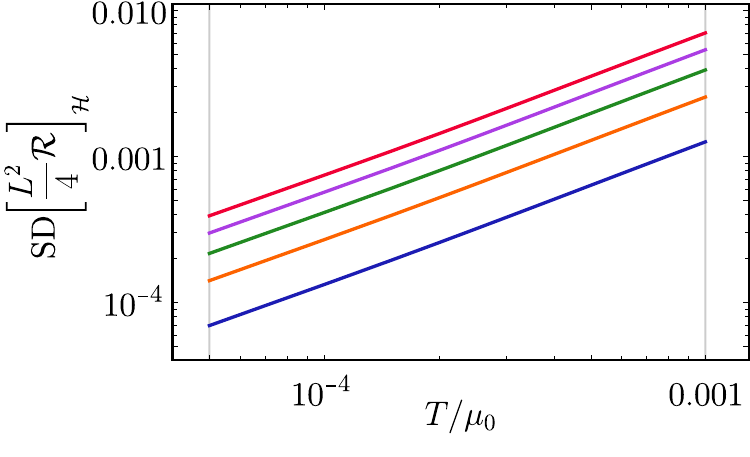}
    \end{subfigure}
    \caption{Left: Ricci scalar integrated over the disordered BTZ horizon for the smallest and largest disorder strengths.
    Both datasets match the clean case (grey dashed line).
   Vertical grey lines mark the boundaries of the disorder regime, while the horizontal grey line shows the zero temperature (AdS$_2\times \mathbb{R}$) value.
    Right: The standard deviation of the Ricci scalar at the horizon is shown for disorder strengths $V/\sqrt{\mu_0} = \{0.2, \ 0.4,\ 0.6, \ 0.8, \ 1 \}$ (from blue to red). It fits nicely to a power law, $ \text{SD}\left[\mathcal{R}\right]_\mathcal{H}\sim T^{0.97}$ for all disorder strengths.}
    \label{fig:AdS3_Ricci_Mean_Std}
\end{figure}

In three dimensions, we lack a direct probe of the horizon geometry analogous to
$\mathcal{W}$ in \eqref{eq:calW} for AdS$_4$.
Our analysis is therefore restricted to characterizing the near-horizon region through the entropy and curvature invariants.
As shown in Fig.~\ref{fig:AdS3Entropy}, the entropy behaves as in the clean case for all disorder strengths.
The same holds for the average of the Ricci scalar along the bifurcation surface, plotted in the left panel of Fig.~\ref{fig:AdS3_Ricci_Mean_Std}.
This behavior was also observed in the four-dimensional scenario. A crucial distinction, however, arises here: the amplitude of disorder-induced perturbations 
decreases as temperature is lowered. 
Indeed, the right panel of Fig.~\ref{fig:AdS3_Ricci_Mean_Std} 
shows that the standard deviation of the Ricci scalar at the horizon
falls as a power law for all disorder strengths.
Therefore, in the deep infrared, disorder becomes effectively irrelevant, and the geometry flows back to the AdS$_2 \times \mathbb{R}$ solution of the clean case.


\subsubsection{IR deformation}
\label{sssec:AdS3IRDef}
Figure~\ref{fig:AdS3_Ricci_Mean_Std} shows that in the low $T$ limit the
near-horizon geometry flows to the clean AdS$_2\times\mathbb{R}$ fixed point. This prompts us to compute, as in Section~\ref{sssec:AdS4IRDef}, the scaling dimension of a modulated deformation $\delta A_t$ of the near-horizon background.
In this case we obtain
\begin{equation}
    \Delta (\Delta -1) = 2+ \hat{k}^2 \, .
\end{equation}
Thus, a small inhomogeneity in the chemical potential is irrelevant for all 
$\hat k$, with $\hat k = 2/H_3 \  ( k/\mu_0)^2$.
Consequently, the perturbation cannot modify the IR fixed point, regardless of the magnitude of the wavevector.
This is consistent with the recovery of the clean fixed point at low temperatures.


\subsection{DC conductivity}

We  finally study the DC electrical conductivity of the disordered
BTZ black branes.
Expressions for the DC transport coefficients in this case were given in Section~\ref{ssec:conductivitiesAdS3}.
Moreover, we showed above that the horizon geometry approaches the clean AdS$_2\times\mathbb{R}$ fixed point as $T\to0$.
Hence we expect translation invariance to be restored in that limit:
momentum conservation is reinstated in the IR, dissipative processes are suppressed, and transport coefficients diverge.

\begin{figure}
    \centering
    \includegraphics[width=0.6\linewidth]{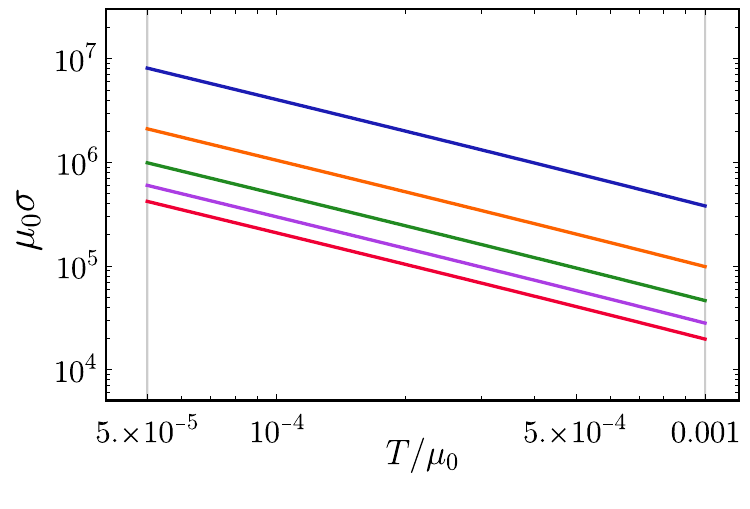}
    \caption{DC electrical conductivity of the disordered BTZ black brane as a function of temperature for several disorder strengths (as in Fig.~\ref{fig:AdS3_Ricci_Mean_Std}).}
    \label{fig:AdS3Cond}
\end{figure}

Fig. \ref{fig:AdS3Cond} displays the DC electrical conductivity $\sigma$.
As temperature is lowered, $\sigma$ grows as a power law.
This behavior is consistent with our observation that the geometry approaches the clean IR fixed point AdS$_2 \times \mathbb{R}$ as $T\to0$.


\section{Conclusions}\label{section:Conclusions}
\label{sec:conclusions}
In this work we have constructed holographic duals of strongly coupled systems with Harris-relevant disorder sourced by a noisy chemical potential in two and three spacetime dimensions at finite temperature and nonzero (mean) chemical potential. 
These backgrounds feature charged and highly inhomogeneous planar horizons in asymptotically AdS$_3$ and AdS$_4$ spacetimes, respectively.
Numerically, disorder is implemented through the spatially-dependent chemical potential
$\mu(x)$ in~\eqref{eq:dismu}.
This function is
a sum of $N$ cosines with wavevectors
$k_n=k_0/n$ and random phases $\delta_n$ (with $n=1,\dots,N$).
The largest wavevector $k_0$ is a UV cutoff of the disordered regime of the system, akin to a minimum distance between impurities.
In the limit $N\to\infty$, $\mu(x)$
reproduces Gaussian white noise;
in our numerical simulations we keep $N$ finite, so the disordered regime
is restricted to temperatures between the UV and IR cutoffs $k_0$ and $k_0/N$.

Our fully backreacted solutions reveal how disorder affects the geometry as one approaches the black hole horizon. In the dual field theory, this corresponds to how the disordered theory flows under the RG and thus probes the interplay between strong interactions and disorder.

In the AdS$_4$ case, disorder affects the horizon of the black hole, rendering it inhomogeneous. The inhomogeneity persists as the temperature is lowered,
showing that disorder modifies the low temperature physics of the theory.
Although this outcome might be expected since the disordered source is Harris-relevant, the effect is milder than one might anticipate.
While disorder survives in the IR and modifies some averaged quantities
(see Fig.~\ref{fig:LowT_AdS4_entropy}),
the averaged geometry is nevertheless the clean one. 
In the IR, we find a continuous family of solutions whose averaged
geometry approaches AdS$_2 \times \mathbb{R}^2$, differing by the norm of the Killing vector orthogonal to the disordered direction; that norm being set by the disorder strength $V$.
This behavior makes disorder resemble a marginal deformation of the clean theory rather than a relevant one.
We confirmed this by computing the conformal dimension of a modulated gauge field perturbation in the near-horizon AdS$_2 \times \mathbb{R}^2$ throat as in~\cite{Hartnoll:2012rj}, finding $\Delta\sim 1+k^4$,
so the perturbation is marginal in the small $k$ limit.

The contradiction with the Harris criterion is starker in the AdS$_3$ case. 
At high temperatures, the black holes are inhomogeneous, but as the temperature is lowered, the inhomogeneities decay and the near-horizon geometry approaches the clean AdS$_2 \times \mathbb{R}$
fixed point (see Fig.~\ref{fig:AdS3_Ricci_Mean_Std}).
Thus, disorder behaves as an irrelevant deformation in this setup.
Here, the conformal dimension of a modulated gauge field perturbation
in the near-horizon AdS$_2 \times \mathbb{R}$ throat is $\Delta\sim 2+k^2$, consistent with the irrelevancy of disorder in the IR geometries.

Our numerical solutions therefore, indicate 
a violation of the Harris criterion in (some) systems with holographic duals. The literature contains examples of Harris-criterion violations in certain random lattice contexts \cite{Schrauth:2018, Topchyan:2024aqw},
but those setups differ from ours, which lack a microscopic field theory description. 
Nonetheless, our findings can be explained in terms of the scaling dimension of modulated perturbations about the IR geometries of the corresponding homogeneous charged black holes.

The picture we have described is supported by our study of DC transport.
We expressed the DC thermoelectric conductivities as integrals over the horizon 
and thus related DC transport to properties of the inhomogeneous horizon geometries
of the gravity dual.
Focusing on the DC electrical conductivity $\sigma$, we find a distinct behavior in the two setups we studied.
In AdS$_4$, $\sigma$ stabilizes within the disordered regime, and our data indicate a finite residual resistivity as $T \to 0$.
This residual resistivity arises from the inhomogeneous nature of the near-horizon geometry at low temperature.
Note that homogeneous AdS$_4$ models with momentum relaxation~\cite{Andrade:2013gsa} also
display finite $T$-independent resistivity.
Although $\sigma$ decreases as disorder strength is increased, it remains above the lower bound proved in \cite{Grozdanov:2015qia} for all temperatures and disorder strengths we studied.
By contrast, in the AdS$_3$ geometries (dual to 1+1 dimensional systems)
the DC electrical conductivity does not stabilize as $T\to0$;
it increases as a power law, consistent with the restoration of the
AdS$_2\times\mathbb{R}$ clean fixed point.

The analysis presented in this work raises interesting questions about the interplay of disorder and strong interactions.
First, does the observed Harris-criterion violation persist in higher dimensions, or is it specific to AdS$_4$ and AdS$_3$? What modified criterion governs the relevance of disorder in strongly interacting systems? 
Could the disorder-averaged infrared physics be captured by an effective theory of momentum relaxation with a homogeneous dual geometry?
What are the effects of disorder on superconducting phase transitions?
Does it suppress $T_c$, or could it potentially enhance superconductivity via mechanisms such as the islands reported in~\cite{Arean:2015sqa}?
These and other key questions could be addressed by pursuing the following directions.
\begin{itemize}
    \item Examine AC transport in the disordered geometries constructed here. It would be valuable to compare 
    these responses with the results of~\cite{Balm:2022bju}, where holographic quantum critical points in a periodic potential displayed features characteristic of strange metals.

\item Compare these explicitly disordered models  with homogeneous momentum-relaxing setups as
~\cite{Vegh:2013sk,Andrade:2013gsa,Baggioli:2014roa,Davison:2014lua,Donos:2014uba}. Such homogeneous backgrounds are considered effective descriptions of impurities in strongly coupled materials. A detailed comparison could clarify 
if those effective models capture the physics of disorder.

\item Study the extremal limit of the disordered system. Building on the approach followed in Appendix~\ref{app:modads4}, it would be interesting to consider a disordered chemical potential in the extremal AdS black hole geometry.

\item Explore the interplay of disorder and spatial ordering, which is relevant for the description of strange metallic
    phases~\cite{Amoretti:2017axe,Baggioli:2022pyb}.
    
\item Analyze superconducting instabilities in disordered backgrounds.
    In particular, determine how disorder affects the critical temperature and the structure of the superconducting phase, including the possible emerging of islands enhancing superconductivity~\cite{Vojta_2019,Arean:2015sqa}; and analyze
    the interplay of disorder and superfluid and spatial
    ordering~\cite{Cremonini:2017usb,Ling:2019gjy,Li:2024ybq}.

\item Investigate disordered chemical potentials about charge neutrality.
    When the clean background is neutral, qualitatively different phenomenology
    is expected~\cite{OKeeffe:2015qma,Garcia-Garcia:2015crx};
    this may be relevant for the dynamics of graphene~\cite{Lucas:2015sya}.

 \item Explore the interior geometry of these highly inhomogeneous black holes and  compute dual observables that probe the region behind the horizon~\cite{Frenkel:2020ysx,Jorstad:2023kmq,Mansoori:2021wxf,Arean:2024pzo}.

\item Study disorder in Einstein-Maxwell-Dilaton models, tuning the dimension of the disordered coupling to search for the disordered fixed points predicted in~\cite{Huang:2023ihu}.
\end{itemize}

Holography provides a unique, non-perturbative framework to address these questions,
thus probing the interplay between disorder and strong correlations in regimes inaccessible to conventional methods.
The disordered geometries constructed here offer a controlled setting to test ideas about disorder-induced fixed points, transport, and emergent IR dynamics.
We expect that further analytic and numerical study of these backgrounds will yield insights relevant both to condensed-matter systems and to the holographic understanding of disordered quantum criticality.

\section*{Acknowledgments}
The work of D.A and P.G.R.
is supported through the grants CEX2020-001007-S and PID2021-123017NB-100, PID2021-127726NB-I00 funded by MCIN/AEI/10.13039/501100011033 and by ERDF ``A way of making Europe''. This work is supported by the U.S. Department of Energy grant DE-FG02-97ER-41014 (UW Nuclear Theory).
S.G. was supported in part by a Feodor Lynen Research Fellowship of the Alexander von Humboldt Foundation.
This work is partially funded by the European Commission – NextGenerationEU, through Momentum CSIC Programme: Develop Your Digital Talent. We acknowledge HPC support by Emilio Ambite, staff hired under the Generation D initiative, promoted by Red.es, an organisation attached to the Spanish Ministry for Digital Transformation and the Civil Service, for the attraction and retention of talent through grants and training contracts, financed by the Recovery, Transformation and Resilience Plan through the EU’s Next Generation funds.
We would like to thank I.~Salazar~Landea for his involvement in early stages of this work, and D.~Garc\'\i a Fariña and B.~Goutéraux for useful discussions.

\appendix

\section{Equations of motion and boundary conditions}
\label{app:eomsbcs}

In this appendix we present the PDEs that describe the inhomogeneous system studied in this work and summarize our numerical procedure.
The equations of motion derived from the action~\eqref{eq:einsmaxwellaction} are
\begin{subequations}
\begin{align}
    & R_{\mu \nu}+ d \  g_{\mu \nu} - \frac{1}{2}\left[F_{\mu \alpha} F_{\nu}^{\ \alpha} - \frac{1}{2(d-1)} F^2 g_{\mu \nu}\right] =0 \, , \\
    & \nabla_\mu F^{\mu \nu} =0 \, .    
\end{align}
\end{subequations}
To construct inhomogeneous solutions
we substitute the ansatzes \eqref{eq:ads4ansatz} for AdS$_4$ and \eqref{eq:ads3ansatz} for AdS$_3$ into these equations of motion.
Those ansatzes result in a consistent set of PDEs but 
do not fix the diffeomorphism invariance along the $(z,x)$ directions.
As in~\cite{Donos:2014cya} we fix the gauge by means of the DeTurck trick~\cite{Headrick:2009pv,Horowitz:2012ky,Dias:2015nua},
which renders the resulting system of PDEs elliptic and yields a well-posed boundary value problem.
One proceeds by modifying Einstein's equations as 
\begin{subequations}
\label{eqs:generaleoms}
\begin{align}
    & R_{\mu \nu} - \nabla_{(\mu} \xi_{\nu)} + d \  g_{\mu \nu} - \frac{1}{2}\left[F_{\mu \alpha} F_{\nu}^{\ \alpha} - \frac{1}{2(d-1)} F^2 g_{\mu \nu}\right] = 0 \, ,
    \label{eq:einsdeturckeqs}\\
    & \nabla_\mu F^{\mu \nu} = 0 \, .    
\end{align}
\end{subequations}
We added the gradient of the DeTurck vector $\xi^\mu = g^{\alpha \beta} (\Gamma^\mu_{\alpha \beta} (g) - \Gamma^\mu_{\alpha \beta} (\bar{g}))$, where $\Gamma^\mu_{\alpha \beta} (\bar{g})$ is the Levi-Civitta connection for a reference metric $\bar{g}$ that shares the same horizon structure as the original metric $g$. For simplicity, we choose $\bar{g}$ to be the homogeneous (clean) geometry at the same temperature. 
The inclusion of the DeTurck term renders the system of PDEs elliptic.
Maxwell's equations can, in principle, suffer similar gauge-related pathologies \cite{Rozali:2013fna}; however in our setup
$A \propto dt$ and the spacetime is static, so we can adopt a gauge in
which Maxwell's equations are well-posed without an additional DeTurck-like term \cite{Dias:2015nua}.
Physical solutions are those with vanishing DeTurck vector, and thus we need to check that indeed $\xi^\mu=0$ for our numerical simulations.
We monitor this by keeping track of $\xi^2$ (which is positive-definite) and ensuring that it decreases as we refine our numerics. 

We  now discuss the boundary conditions
we need to impose on the boundary of AdS and at the horizon of the disordered black branes.


\subsection{Disordered AdS$_4$-RN}
\label{app:eomsbcsads4}

Substituting the AdS$_4$ ansatz \eqref{eq:ads4ansatz} into the
equations~\eqref{eqs:generaleoms} 
yields six coupled second order PDEs.
The boundary conditions at the UV boundary $z=0$ were discussed around eq.~\eqref{eq:uvbcsads4}.
They require the metric to approach  AdS$_4$
and the temporal component of the gauge field to approach the disordered profile
$\mu(x)$ of \eqref{eq:dismu}.
The IR boundary conditions follow from
enforcing regularity at the horizon (at $ z = 1$)
and applying the zeroth law of black hole thermodynamics, which requires the surface gravity to be constant. This implies
\begin{subequations}
\label{eq:AdS4IRp1}
\begin{align}
&A_t(1,x)= 0  \, ,\\
&h_2(1, x)=h_1(1,x) \, .
\end{align}
\end{subequations} 
For the remaining four IR boundary conditions we require regular Taylor expansions
at the horizon. Concretely, for each field 
$\mathcal{F} =  \left\{A_t, h_3, h_4, h_5 \right\}$ we set
\begin{equation}
 \mathcal{F}(z, x) = \mathcal{F}_0(x) + (1-z) \mathcal{F}_1(x)  + ... \, ,
\end{equation}
which, together with~\eqref{eq:AdS4IRp1}, enforce regularity and provide a well-posed set of boundary conditions compatible with $\xi^2 =0$ at the horizon.


\subsection{Disordered AdS$_3$-BTZ}
\label{app:eomsbcsads3}

For the ansatz~\eqref{eq:ads3ansatz}, the field equations~\eqref{eqs:generaleoms}
reduce to five coupled second order PDEs.
The UV conditions for the metric functions at the AdS boundary are analogous to the 
four-dimensional 
AdS$_4$ case and are given in~\eqref{eq:ads3metricbc}.
For the gauge field we require the temporal component to behave toward $z=0$ as
\begin{equation}
    A_t = \mu(x) \log(z) + \rho(x) + ...\,,
\end{equation}
with $\mu(x)$ the disorder profile~\eqref{eq:dismu} and
the ellipsis denoting terms that vanish as $z\to0$.
This singular behavior of $A_t$ is challenging for numerical simulations. Thus we subtract the asymptotic logarithmic contribution and solve instead for 
\begin{equation}
\tilde A_t(z,x) = z\left(A_t - \mu(x)\log(z)\right),
\end{equation}
which must vanish at the AdS boundary for our solutions.
Thus we impose the UV boundary condition $\tilde A_t(0,x)=0$.

As in the AdS$_4$ case above, the IR boundary conditions we impose
are such that the horizon is regular and the zeroth law is not violated
\begin{equation}
\begin{aligned}
A_t(1, x)  &= 0 \,,  \\
H_1(1, x) &= H_2(1, x) \,,   \\ 
\mathcal{F}(z, x) &= \mathcal{F}_0(x) + (1-z) \mathcal{F}_1(x)  +... \, ,
\end{aligned}
\end{equation}
with $\mathcal{F} = \left\{A_t, H_3, H_4\right\}$. Again, these are enough to impose $\left. \xi^2\right|_\mathcal{H}=0$.


\section{Modulated AdS$_4$ black branes}
\label{app:modads4}

Although outside of the scope of this paper, 
it is compelling to study extremal AdS$_4$ black branes with a spatially modulated charge density. 
In light of our results, it is particularly interesting to ask whether
the horizon inhomogeneities we observed
persist in the strict $T=0$ limit for wavevectors $k$ in the range explored in this work.

One might expect any finite-$k$ modulation to ebb away in 
the IR, since its scaling dimension in the AdS$_2\times \mathbb{R}^2$ throat
is always irrelevant, $\Delta\sim 1+k^4$ (see eq.~\eqref{eq:IRdimAdS4}).
However, for disordered systems, probing the disordered regime
requires keeping the temperature above the smallest $k$ in the disorder distribution
(see the discussion around eq.~\eqref{eq:discutoffs}), which
would lead us to consider modulations with $k\to0$ as $T\to0$.
It is then reasonable to expect that, as seen in~\cite{Donos:2014yya} for relatively large $k$, modulations with extremely small wavevector result in inhomogeneities (of the horizon) that die as $T\to0$ extremely slowly. 
Alternatively, as observed in \cite{Hartnoll_2014flopy}, extremal
inhomogeneous charged horizons may exist for sufficiently small $k$.

In Section~\ref{ssec:ads4lowt} we  observed that the
inhomogeneities of the near-horizon geometry survive in the low temperature limit.
To support those findings,
here we construct modulated AdS$_4$ charged black branes at low and zero temperature.
Instead of a disordered chemical potential as~\eqref{eq:dismu}, we now consider a single cosine, namely
\begin{equation}
\mu(x) = \mu_0\left(1+w\cos(k\,x)\right).
\label{eq:modmu}
\end{equation}
Our solutions are parametrized by 
the dimensionless ratio $k/\mu_0$ and the amplitude $w$.
As in the disordered case, we employ the ansatz~\eqref{eq:ads4ansatz} and, applying the same numerical methods, construct solutions with $k/\mu_0$ in the range relevant for the disorder realizations explored in Section~\ref{sec:ads4}.

\begin{figure}
    \centering
 \begin{subfigure}[b]{0.50\linewidth}
            \centering
            \includegraphics[width=\linewidth]{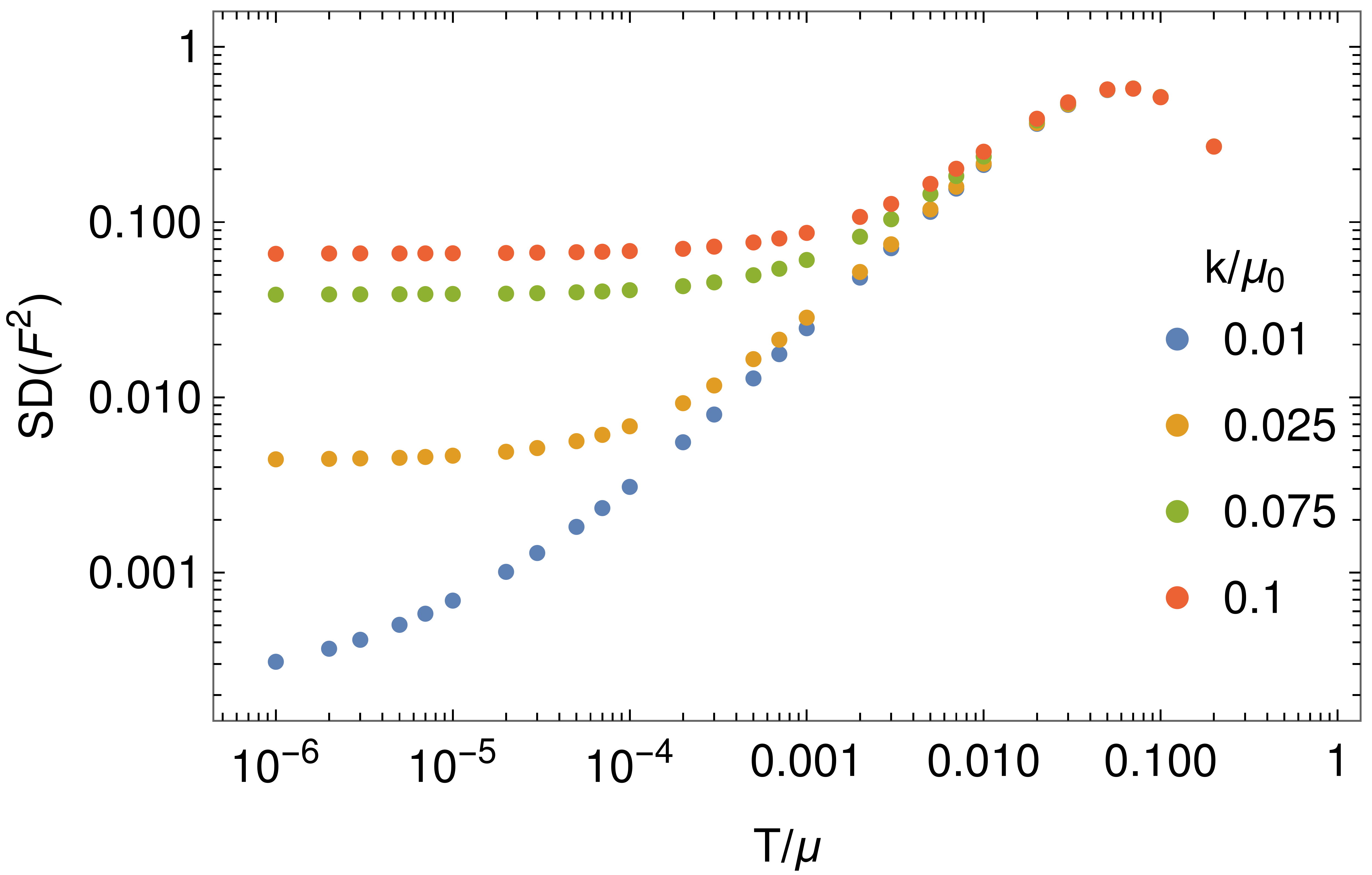}
            \captionsetup{justification=centering}
        \end{subfigure}
        \begin{subfigure}[b]{0.475\linewidth}
            \centering
            \includegraphics[width=\linewidth]{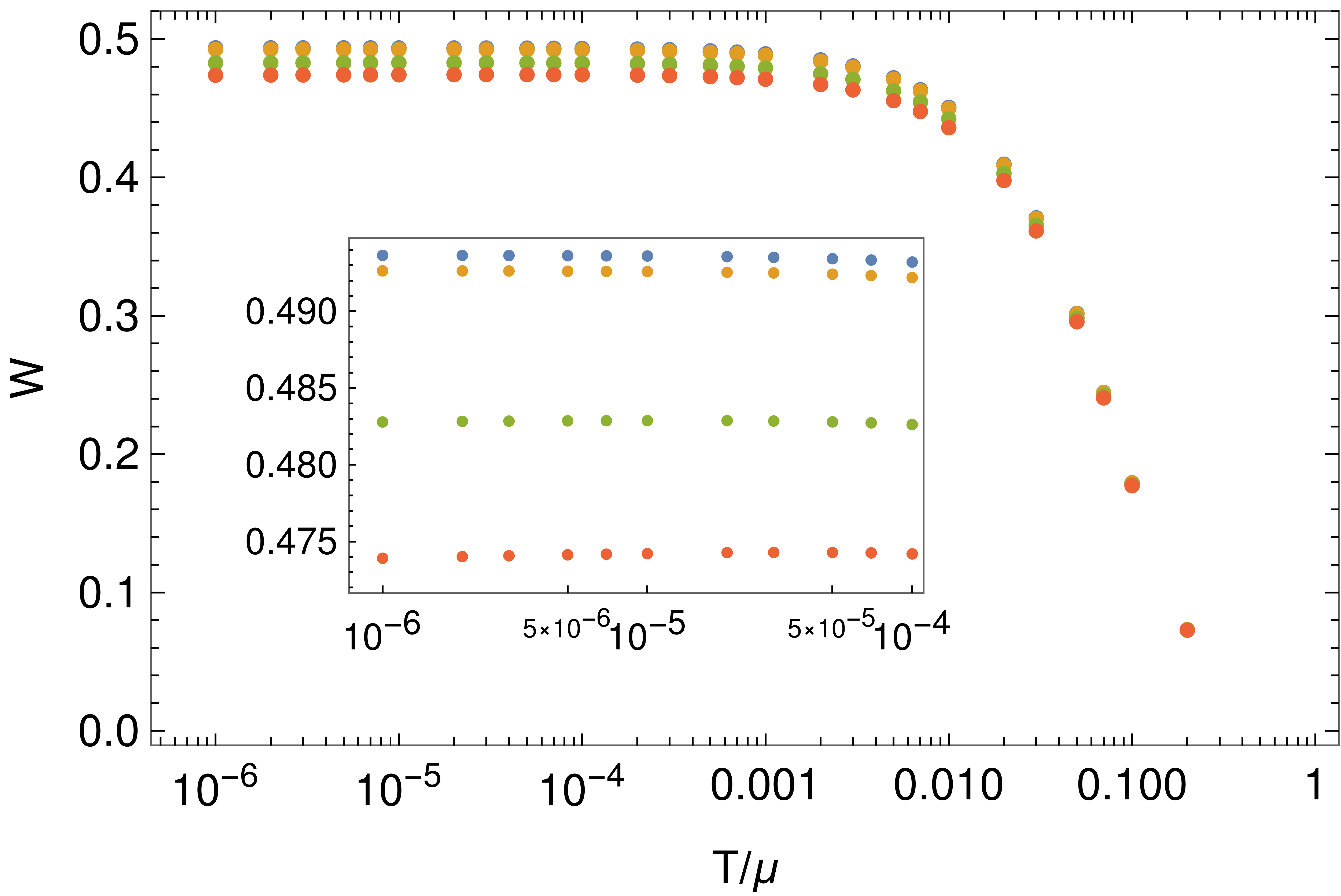}
        \end{subfigure}
    \caption{Left: Standard deviation of $F^2$ as a function of temperature for modulated AdS$_4$ black branes with $w=0.1$ and different values of $k/\mu_0$. Right: corresponding values of $W$
    for the solutions shown in the left panel.}
    \label{fig:AdS4_modhor}
\end{figure}

Our main goal is to characterize the low temperature near-horizon geometry of the modulated black branes,
and in particular
to determine whether the horizon remains inhomogeneous as $T\to0$.
To that end, in the right panel of Fig.~\ref{fig:AdS4_modhor} we  plot the quantity
\begin{equation}
W={\rm max}(g_{yy})/{\rm min}(g_{yy})\big|_h-1\,,
\label{eq:curlyw}
\end{equation}
which measures the horizon modulation along the transverse direction $y$.
For $k/\mu_0\leq 0.1$,
$W$ approaches  a constant as $T\to0$. We confirm this below by constructing the corresponding zero temperature solution.
In the left panel of Fig.~\ref{fig:AdS4_modhor} we also show the standard deviation of $F^2$ at the horizon, which, similarly to the disordered case in Fig.~\ref{fig:AdS4_modhor},
tends to a nonzero value as $T\to0$.
We verify this in the zero temperature solutions below.


\subsection{Zero temperature}
To further confirm the modulated nature of the horizon, we now construct zero temperature modulated black branes for representative values of $k_0/\mu$ in the range explored above. Following~\cite{Hartnoll_2014flopy}, we consider the ansatz
\begin{align}
\label{eq:ads4T0ansatz}
&ds^2= {1\over z^2}\left[
-h_1\,G\,(1-z)^2\,dt^2+{h_2\over G\,(1-z)^2}\,(dz+h_4\,dx)^2+h_3\,dx^2
+h_5\,dy^2\right]\,,\\
&A = A_t(z,x)\,dt\,,\qquad {\rm and}\quad G=1+2z+3z^2\,.
\end{align}
As above, $h_1$, $h_2$, $h_3$, $h_4$, $h_5$ and $A_t$ are functions of $z$, and $x$.
The extremal homogeneous Reissner-Nordstr\"om black brane is recovered for $h_1=h_2=h_3=h_5=1$, $h_4=0$, and $A_t=\sqrt{12}(1-z)$.
We generate modulated extremal black brane solutions by imposing the boundary conditions
\begin{equation}
A_t(0,x)=2\sqrt{3}\left(1+w\cos(k\,x)\right)\,,\quad
h_1(0,x)=h_2(0,x)=h_3(0,x)=h_5(0,x)=1\,,\;h_4(0,x)=0\,,
\end{equation}
which correspond to a baseline chemical potential of the extremal solution $\mu_0=2\sqrt{3}$. 
The resulting family of solutions is parametrized by $k/\mu_0$ $w$.

\begin{figure}
    \centering
 \begin{subfigure}[b]{\linewidth}
            \centering
            \includegraphics[width=0.49\linewidth]{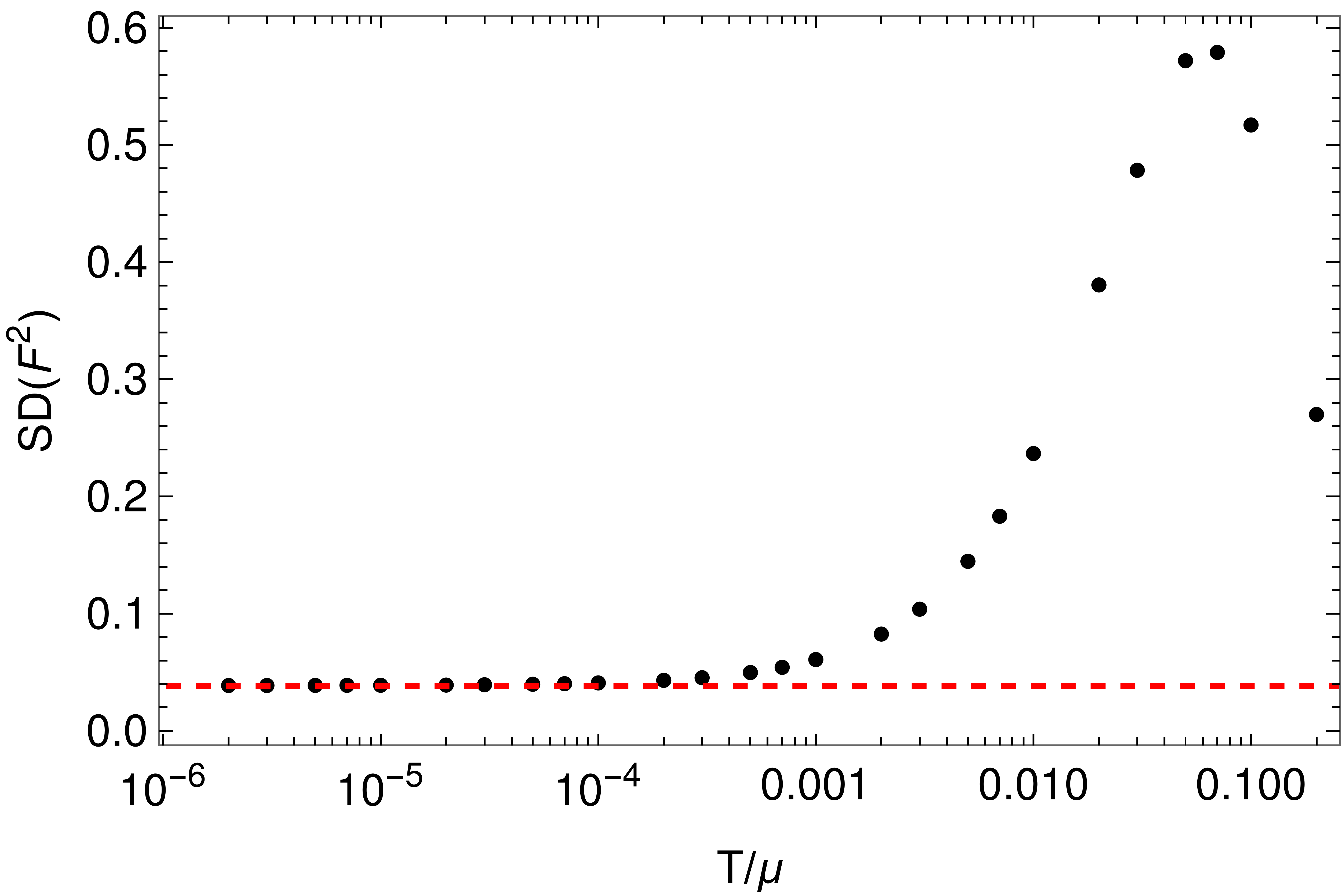}
            \includegraphics[width=0.49\linewidth]{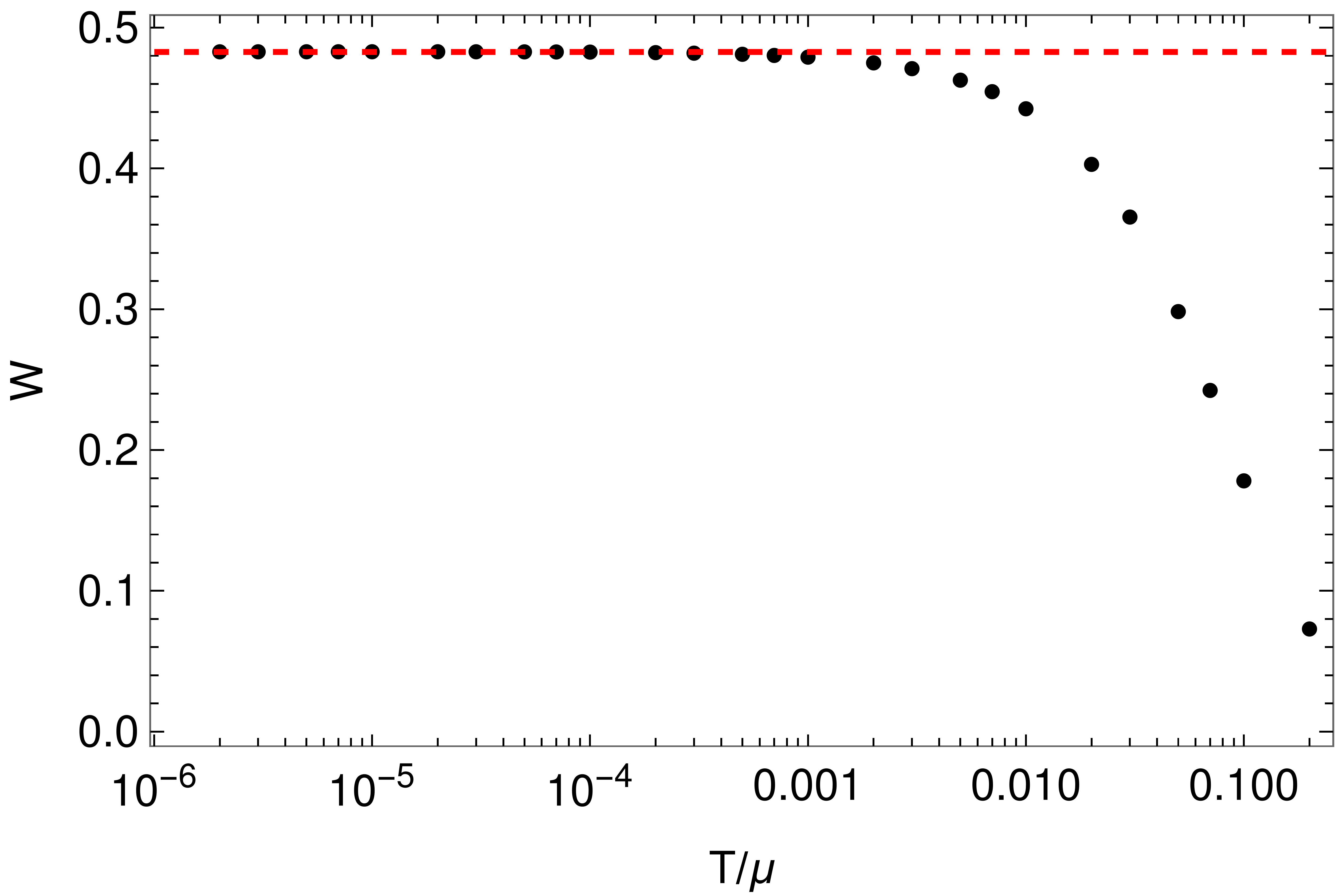}
            \captionsetup{justification=centering}
            \caption{$k/\mu_0=0.075$}
        \end{subfigure}    
\begin{subfigure}[b]{\linewidth}
            \centering
            \includegraphics[width=0.49\linewidth]{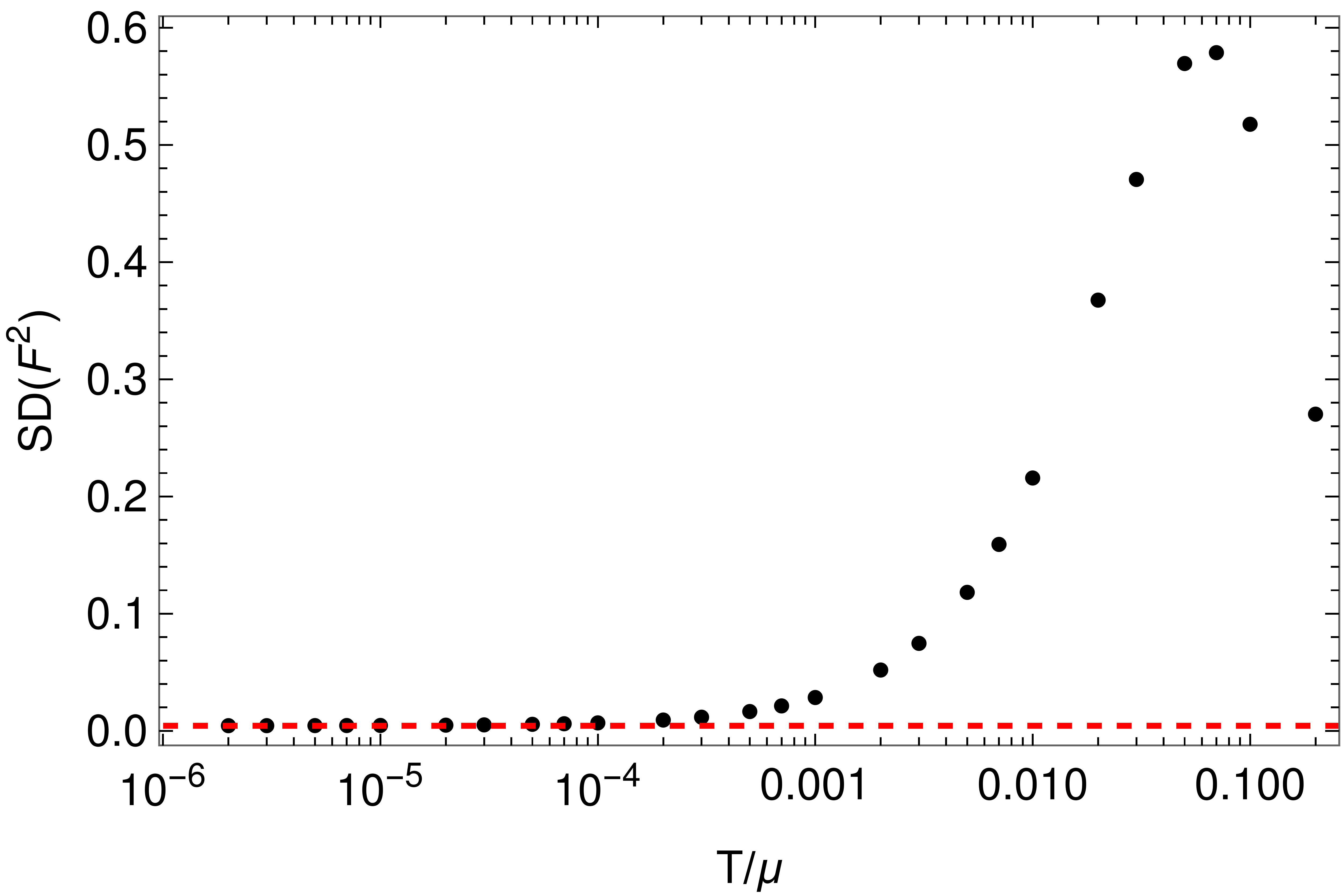}
            \includegraphics[width=0.49\linewidth]{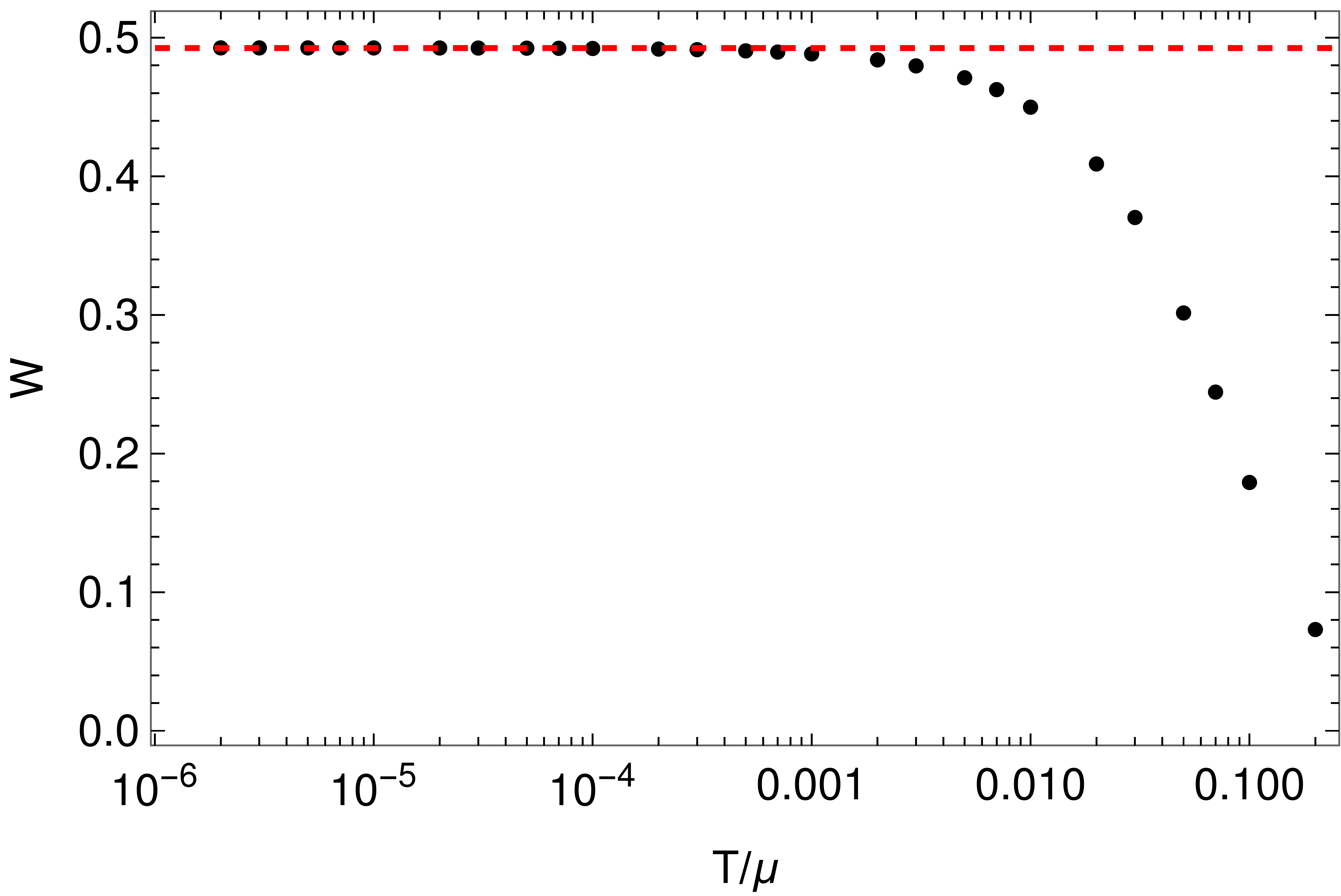}
            \captionsetup{justification=centering}
            \caption{$k/\mu_0=0.025$}
        \end{subfigure}
    \caption{Left: standard deviation of $F^2$ at the horizon as a function of temperature for $w=0.1$ and two values of $k/\mu_0$ (as indicated).
    The red dashed line shows the value obtained at zero temperature. Right: $W$ versus temperature for the same parameters, with the red dashed line again denoting the zero-temperature value.}
    \label{fig:AdS4_T0_crlw}
\end{figure}

\begin{figure}
    \centering
 \begin{subfigure}[b]{\linewidth}
            \centering
            \includegraphics[width=0.48\linewidth]{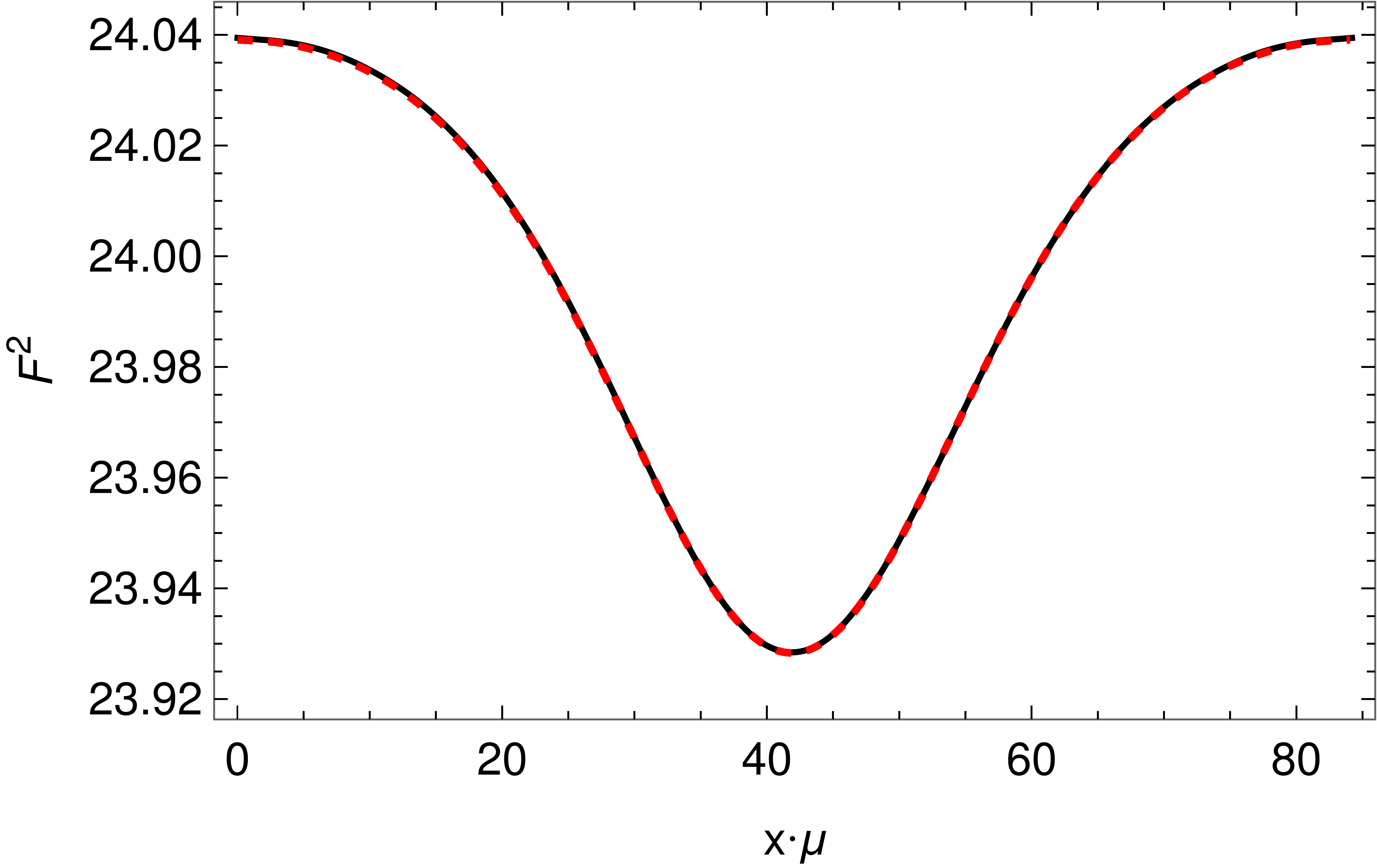}
            \includegraphics[width=0.49\linewidth]{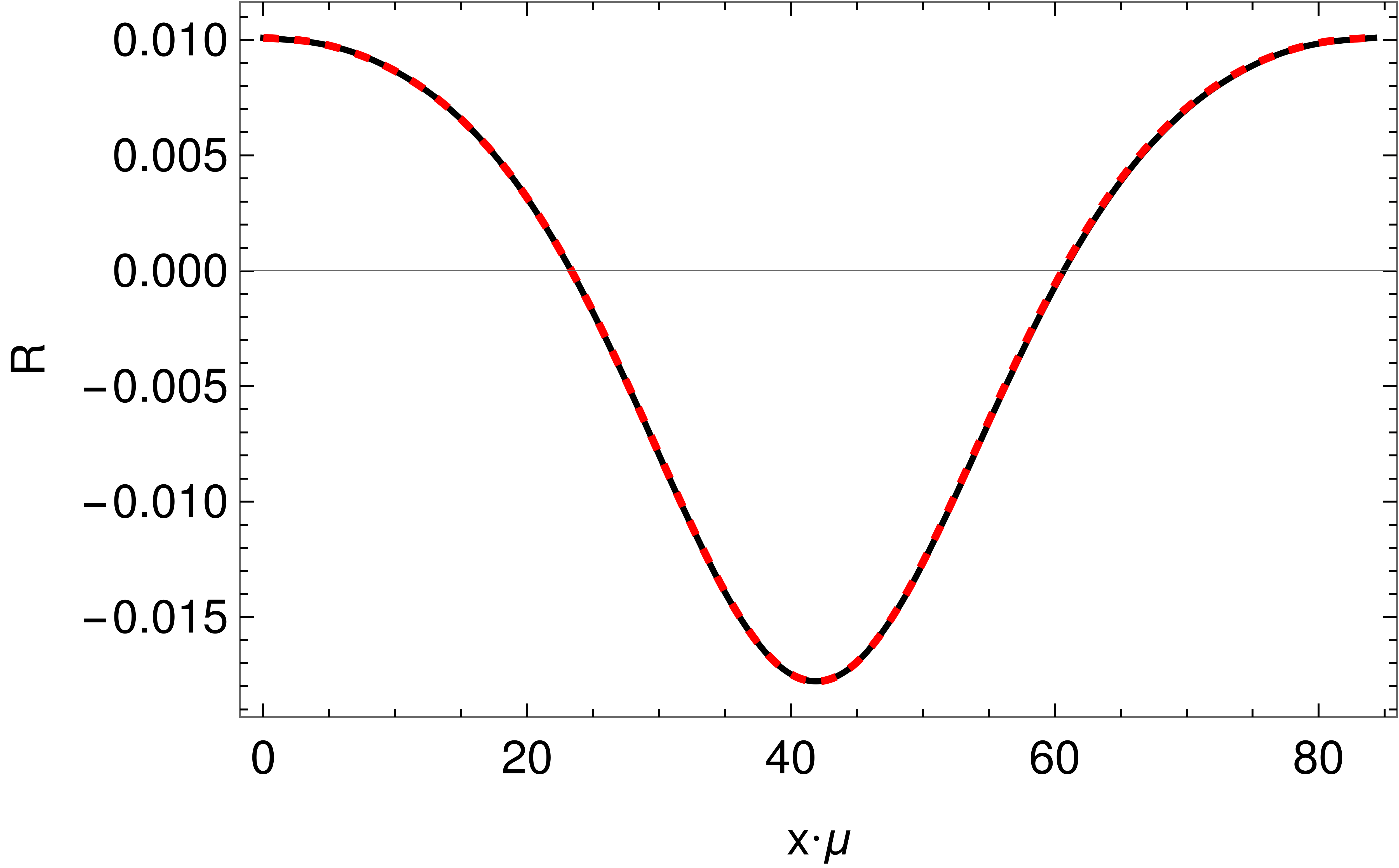}
            \caption{$k/\mu=0.075$}
        \end{subfigure}
        \begin{subfigure}[b]{\linewidth}
            \centering
            \includegraphics[width=0.48\linewidth]{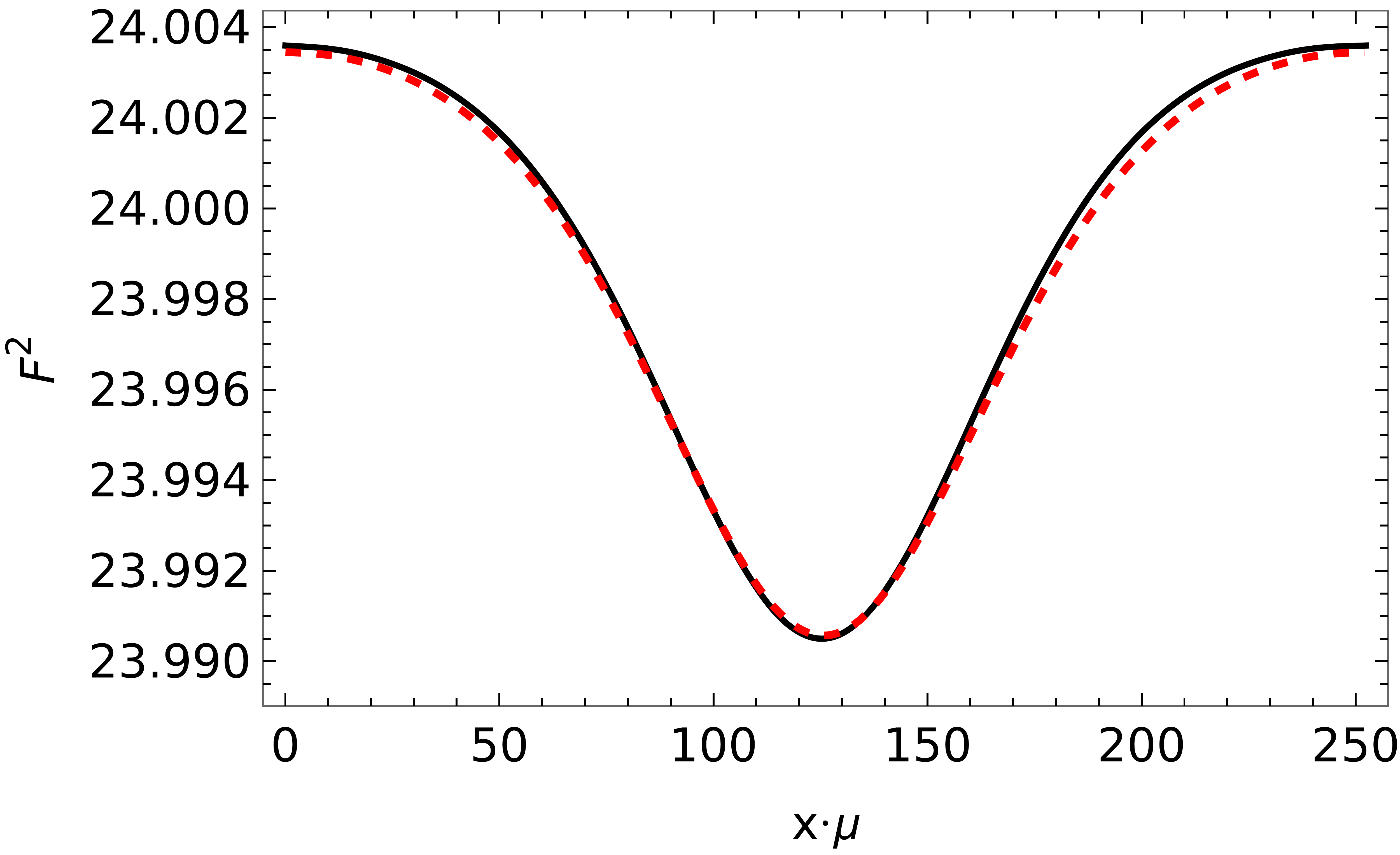}
            \includegraphics[width=0.49\linewidth]{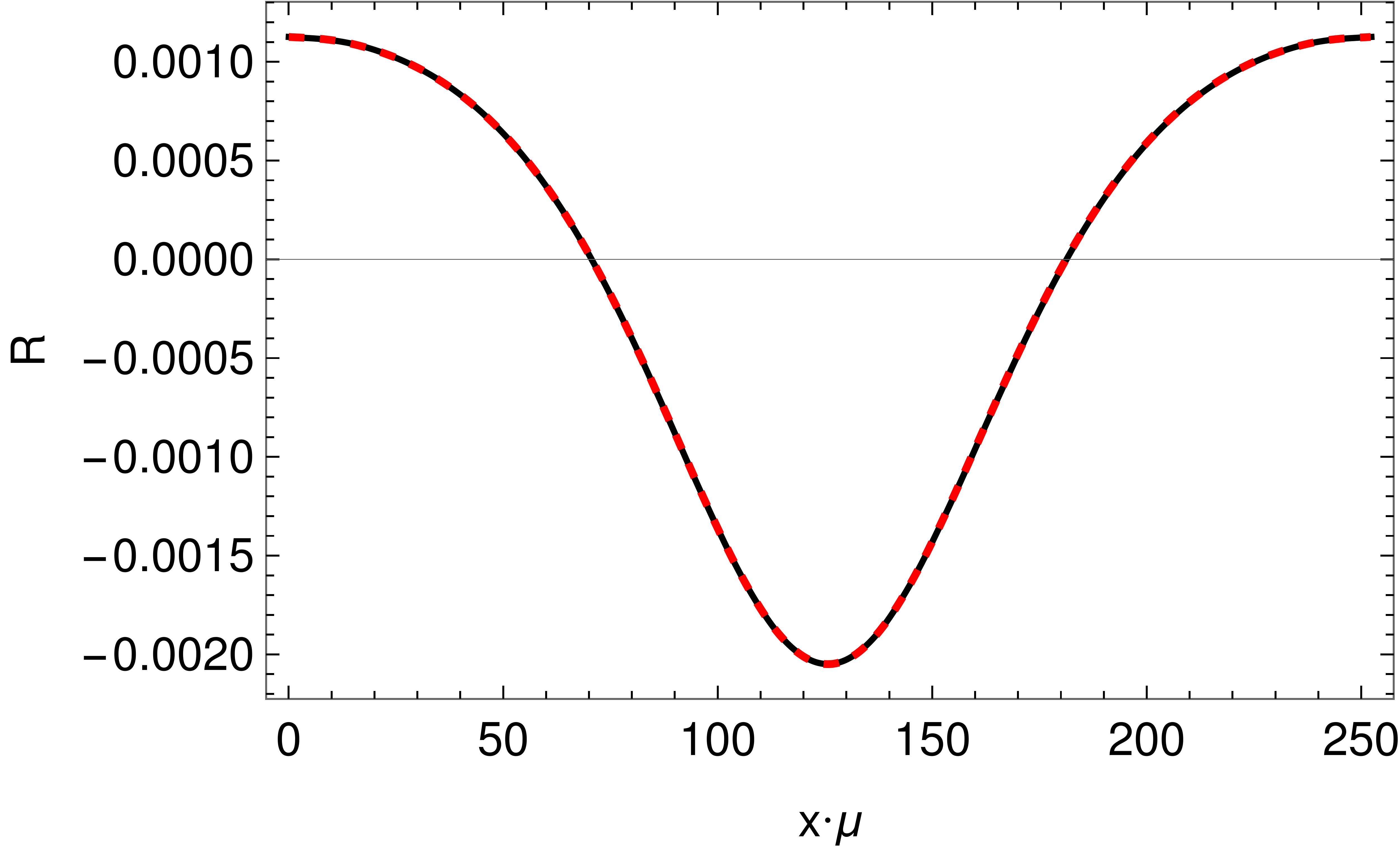}
            \caption{$k/\mu=0.025$}
        \end{subfigure}
    \caption{Left: $F^2$ at the horizon for modulated black branes with $w=0.1$ and $k/\mu=0.075$ (a) and $k/\mu=0.025$ (b) . The black line corresponds to the zero temperature solution and the red dashed line to $T/\mu=3\times10^{-6}$. Right: induced Ricci scalar at the horizon for the same solution as in the left panel.}
    \label{fig:AdS4_T0_hordata}
\end{figure}

In Fig.~\ref{fig:AdS4_T0_crlw} we plot ${\rm SD}(F^2)$ and $W$ versus temperature for modulated black branes with $k/\mu_0=0.075$ and
$k/\mu_0=0.025$, superimposing in red the $T=0$ value.
Figure~\ref{fig:AdS4_T0_hordata} displays $F^2$ and the Ricci scalar at the horizon for zero temperature and very low temperature. 
These figures show that the $T\to0$ limit of the modulated brane converges to the zero temperature modulated horizon.

Next, in Fig.~\ref{fig:AdS4_T0_hordata} we show $F^2$ and the Ricci scalar at the horizon for zero temperature and very low temperature. 
These figures make clear that the low $T$ limit of the modulated brane converges to the zero temperature modulated horizon.
One can conclude that for wavenumbers in the range examined in~Fig.\ref{fig:AdS4_modhor} the horizon of the modulated extremal black branes is inhomogeneous.

\section{Numerics}
\label{app:num}

In this appendix we provide additional details and consistency checks of our numerical simulations.
We first describe the coordinate maps used to improve numerical accuracy, and then present quality checks of the resulting solutions.

As explained in Appendix~\ref{app:eomsbcs},
we solve a set of coupled PDEs derived from the Maxwell-Einstein action~\eqref{eq:einsmaxwellaction}.
To obtain numerical solutions, we discretize the system using a Chebyshev grid in the
radial direction and a Fourier grid in the spatial direction $x$,
with periodicity $L=2\pi N/k_0$. 
Here $N$ denotes the number of modes in the disordered chemical potential \eqref{eq:dismu}.
The resulting equations are solved using a Newton-Raphson method. All numerical computations were performed in \textit{Mathematica}.


\subsection{Coordinate maps}
To improve the stability of our $\text{AdS}_4$ numerical simulations at low temperatures, where non-analyticities typically develop in the near-horizon region, we implement the coordinate transformation:
\begin{equation}
z = 1 - (1 - \hat{z})^2\,.
\label{eq:zmapads4}
\end{equation}
This map preserves the AdS boundary at $\hat z=0$ and horizon at $\hat z=1$, while
clustering grid points near the horizon.
The quadratic clustering enhances radial resolution in the infrared, 
enabling stable convergence of the Newton-Raphson solver at very low temperatures and helping to keep the DeTurck norm small.

In $\text{AdS}_3$, additional care is required as non-analyticities appear both in the near-horizon region at low temperature, and near the boundary due to the logarithmic divergence of the gauge field. Following \cite{BOYD198949, Grieninger:2020wsb} we use the map
\begin{equation}
    z = \sin^2\left(\frac{\pi}{2}\,\hat{z}\right).
\end{equation}
This coordinate choice has several advantages. Near the horizon ($\hat{z}\to1$) it reduces to $z -1 \propto (1-\hat{z})^2$,
reproducing 
the $\text{AdS}_4$ map \eqref{eq:zmapads4}.
Near the boundary it behaves as $z \propto\hat{z}^2$, 
enhancing UV resolution, while keeping the boundary at $\hat z=0$ and the horizon $\hat z=1$. The combined effect improves the resolution of both infrared and ultraviolet non-analyticities in the low-temperature solutions, thereby maintaining numerical accuracy.


\subsection{DeTurck checks}

As explained in Appendix \ref{app:eomsbcs}, our simulations solve the Einstein-DeTurck equations~\eqref{eq:einsdeturckeqs} rather than the Einstein equations directly. These differ by terms involving the
gradient of the DeTurck vector $\xi$.
In certain cases 
it can be shown
that any solution of the Einstein-DeTurck equations 
with $\xi = 0$ on the boundary
is also a solution of the Einstein equations~\cite{Dias:2015nua}.
However, this
has not been proven in full generality, so it is important to verify that the DeTurck term does not significantly affect our results.

\begin{figure}
    \centering
    \includegraphics[width=0.7\linewidth]{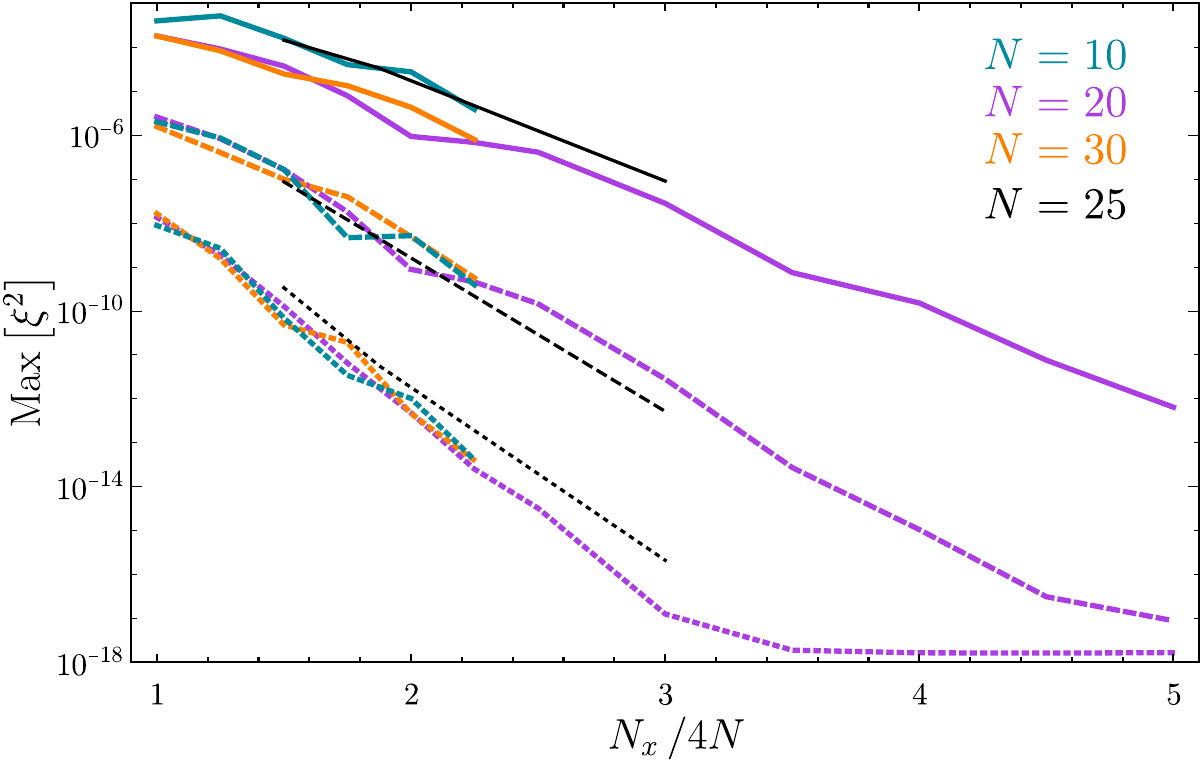}
    \caption{Maximum value of the norm of the DeTurck vector as a function of the number of grid points in the spatial direction, 
    $N_x$, normalized by four times the number of modes, $N$. 
    We present data for different number of modes in the
    disordered chemical potential, as indicated in the legend.
    Dotted, dashed, and solid lines correspond to noise strength
    $V/\sqrt{\mu}=$
    $0.38$, $0.63$, and $1.25$, respectively.
    Black lines are data for the AdS$_4$ setup, while the rest correspond to
    the AdS$_3$ case.
   The line with the lowest noise strength saturates around $10^{-18}$, which approximately reflects the numerical precision of our code. The number of points along the radial direction is fixed to $N_z = 20$ in AdS$_3$ and $N_z = 40$ in AdS$_4$ (black).}
    \label{fig:max_deTurk}
\end{figure}

We asses the quality of our numerics in Fig.~\ref{fig:max_deTurk} by analyzing the dependence of the norm of the DeTurck vector on the number of grid points in the spatial direction $x$, denoted $N_x$.
Increasing the ratio $N_x/N$, where $N$ is the number of modes in the disordered chemical potential~\eqref{eq:dismu}, leads to an exponential decrease of the norm of the DeTurck vector, independent of the 
number of modes.
Our solutions show no qualitative change along the range of $N_x$ shown in the figure.
In the most demanding case,
20 modes at the highest disorder strength,
the averaged Ricci scalar changes by only
$\sim 0.001$ \%,  and the entropy by $\sim 0.05$ \%,
when increasing resolution from 
$N_x / 4 N = 1$ to $N_x / 4 N = 5$,
while the norm of the DeTurck vector decreases by 8 orders of magnitude. 
This indicates that our solutions are, to high numerical precision, genuine solutions of the original Einstein-Maxwell equations. 
In all  plots shown in the main text, we set $N_x / 4 N  \approx 2 $ and require $\xi^2 < 10^{-4}$.

\bibliographystyle{JHEP}
\bibliography{refsdis.bib}
\end{document}